\documentclass[12pt,epsfig,color]{article}
\usepackage{amsmath}
\usepackage{epsfig}
\usepackage{amssymb}
\usepackage{hyperref}
\input{epsf}
\def\beq{\begin{equation}}
\def\eeq{\end{equation}}
\def\beqa{\begin{eqnarray}}
\def\eeqa{\end{eqnarray}}

\newcommand\as{\alpha_s}
\newcommand\f[2]{\frac{#1}{#2}}

\def\beq{\begin{equation}}
\def\eeq{\end{equation}}
\def\beeq{\begin{eqnarray}}
\def\eeeq{\end{eqnarray}}
\def\to{\rightarrow}
\def\nn{\nonumber}

\def\msbar{{\overline {\rm MS}}}

\def\b0{b_0}

\def\th{\hat{\tau}}

\begin{document}

\begin{titlepage}
\renewcommand{\thefootnote}{\fnsymbol{footnote}}
\begin{flushright}
ITP-SB-09-14 \\
%hep-ph/???
     \end{flushright}
\par \vspace{10mm}
\begin{center}
{\large \bf
Threshold Resummation for Di-hadron Production \\[4mm] 
in Hadronic Collisions}
\end{center}

\par \vspace{2mm}
\begin{center}
{\bf Leandro G. Almeida${}^{\,a}$,}
\hskip .2cm
{\bf George Sterman${}^{\,a}$,}
\hskip .2cm
{\bf Werner Vogelsang${}^{\,b}$  }\\[5mm]
\vspace{5mm}
${}^{a}\,$C.N.\ Yang Institute for Theoretical Physics,
Stony Brook University,
Stony Brook, New York 11794 -- 3840, U.S.A.\\[2mm]
${}^{b}\,$Physics Department, Brookhaven National Laboratory, 
Upton, NY 11973, U.S.A.\\
\end{center}

%%%%%%%%%%%%%%%%%%%%%%%%%%%%%%%%%%%%%%%%%%%%%%%%%%%%%%%%%%%%%%%%%%%%%%%%%%%%
%%%%%%%%%%%%%%%%%%%%%%%%%%%       ABSTRACT      %%%%%%%%%%%%%%%%%%%%%%%%%%%%
%%%%%%%%%%%%%%%%%%%%%%%%%%%%%%%%%%%%%%%%%%%%%%%%%%%%%%%%%%%%%%%%%%%%%%%%%%%%

 \vspace{9mm}
\begin{center} {\large \bf Abstract} \end{center}
We study the resummation of large logarithmic perturbative corrections
to the partonic cross sections relevant for di-hadron production
in hadronic collisions, $H_1 H_2\to h_1 h_2 X$, at high 
invariant mass of the produced hadron pair. 
These corrections arise near the threshold for the partonic 
reaction and are associated with soft-gluon emission. We perform
the resummation to next-to-leading logarithmic accuracy,
and show how to incorporate consistently cuts in rapidity
and transverse momentum of the observed particles.
We present numerical results for 
fixed-target and ISR regimes
and find enhancements over the next-to-leading order 
cross section, which significantly improve the
agreement between theoretical predictions and data. 

\end{titlepage}  

\setcounter{footnote}{1}
\renewcommand{\thefootnote}{\fnsymbol{footnote}}

%%%%%%%%%%%%%%%%%%%%%%%%%%%%%%%%%%%%%%%%%%%%%%%%%%%%%%%%%%%%%%%%%%%%%%%%%%%
%%%%%%%%%%%%%%%%%%%%%%%%%%%    INTRODUCTION     %%%%%%%%%%%%%%%%%%%%%%%%%%%
%%%%%%%%%%%%%%%%%%%%%%%%%%%%%%%%%%%%%%%%%%%%%%%%%%%%%%%%%%%%%%%%%%%%%%%%%%%

\section{Introduction}

Cross sections for hadron production in hadronic collisions 
play an important role in QCD. They offer a variety of 
insights into strong interaction dynamics. At sufficiently large 
momentum transfer in the reaction, QCD perturbation theory can be 
used to derive predictions. The cross section may be factorized
at leading power in the hard scale into convolutions of long-distance 
factors representing the structure of the initial hadrons and the 
fragmentation of 
the final-state partons into the observed hadrons, 
and parts that are short-distance and describe the hard 
interactions of the partons. If the parton distribution functions and 
fragmentation functions are known from other processes, especially 
deeply-inelastic scattering and $e^+ e^-$ annihilation, 
hadron production in hadronic collisions directly tests the 
factorized perturbative-QCD approach and the relevance of higher 
orders in the perturbative expansion. 

Much emphasis in both theory and experiment has been on 
single-inclusive hadron production, $H_1 H_2\to 
h X$~\cite{aur,apan,boursoff,kkp1,ddfwv,phenix,star,brahms}. Here 
the large momentum transfer is provided by the high transverse 
momentum of the observed hadron. Of equal importance,
albeit explored to a somewhat lesser extent, is di-hadron production,
$H_1 H_2\to  h_1 h_2 X$, when the pair is produced with large
invariant mass $M$. In many ways, one may think of this process
as a generalization of the Drell-Yan process to a completely
hadronic situation, with the Drell-Yan lepton pair replaced by the hadron
pair. The process is therefore particularly interesting for
studying QCD dynamics, as we shall also see throughout this
paper. Experimental data for di-hadron production as a function
of pair mass are available from various fixed-target 
experiments~\cite{na24,e711,e706}, as well as from the ISR~\cite{ccor}.  
On the theory side, next-to-leading order (NLO) calculations 
for this process are
available~\cite{Chiappetta:1996wp,Owens:2001rr,Binoth:2002wa}. 
They have been confronted with the available data sets, and
it was found that overall agreement could only be achieved
when rather small renormalization and factorizations scales
were chosen. The NLO calculations in fact show very large 
scale dependence. If more natural scales are chosen, NLO theory 
significantly underpredicts the cross section data, as we shall 
also confirm below. 

In the present paper, we investigate the all-order resummation of 
large logarithmic corrections to the partonic cross sections.
This is of considerable interest for the comparison between 
data and the NLO calculation just described. A related resummation 
for the single-inclusive hadron cross section~\cite{ddfwv} was found 
to lead to significant enhancements of the predicted cross
section over NLO, in much better overall agreement with the available
data in that case.

At partonic threshold, when the initial partons have just enough 
energy to produce two partons with high invariant pair mass
(which subsequently fragment into the observed hadron pair), the 
phase space available for gluon bremsstrahlung vanishes, resulting 
in large logarithmic corrections. To be more specific, if we consider the 
cross section as a function of the partonic pair mass $\hat{m}$, 
the partonic threshold is reached when $\hat{s}=\hat{m}^2$, 
that is, $\hat\tau\equiv\hat{m}^2/\hat{s}=1$, where $\sqrt{\hat{s}}$
is the partonic center-of-mass system (c.m.s.) energy.
The leading large contributions near threshold arise as $\as^k
\left[ \ln^{2k-1}(1-\hat\tau)/(1-\hat\tau)\right]_+$ at the $k$th 
order in perturbation theory, where $\as$ is the strong coupling and
the ``plus'' distribution will be defined below. Sufficiently
close to threshold, the perturbative series will be useful only if
such terms are taken into account to all orders in $\as$, which is
what is achieved by threshold resummation~\cite{dyresum,KS,BCMN,dyresum2}. 
Here we extend threshold resummation further, to cross sections
involving cuts on individual hadron $p_T$ and the rapidity of the pair.

We note that this behavior near threshold is very familiar 
from that in the Drell-Yan process, if one thinks of $\hat{m}$
as the invariant mass of the lepton pair. Hadron pair production
is more complex in that gluon emission will occur not only from 
initial-state partons, but also from 
those in the final state.  Furthermore, interference
between soft emissions from the various
external legs is sensitive to the color exchange in the hard 
scattering, which gives rise to a special additional contribution
to the resummation formula, derived in~\cite{KS,KOS,KO1}.

The larger $\hat\tau$, the more dominant the threshold logarithms 
will be. Because of this and the rapid fall-off of the parton
distributions and fragmentation functions with momentum fraction,
threshold effects tend to become more and more relevant as the 
{\it hadronic} scaling variable $\tau\equiv M^2/S$ goes to one. 
This means that the fixed-target regime is the place where threshold
resummation is expected to be particularly relevant and useful. 
We will indeed confirm this in our study. Nonetheless, because
of the convolution form of the partonic cross sections and the
parton distributions and fragmentation functions (see below),
the threshold regime $\hat{\tau}\to 1$ plays an important
role also at higher (collider) energies. Here one may,
however, also have to incorporate higher-order terms that are subleading
at partonic threshold. 

In Sec.~\ref{sec2} we provide the basic formulas for the 
di-hadron cross section as a function of pair mass 
at fixed order in perturbation theory, and display the 
role of the threshold region. Section~\ref{sec3} presents 
details of the threshold resummation for the cross section.  
In Sec.~\ref{sec4} we give phenomenological results, 
comparing the threshold resummed calculation to the 
available experimental data. Finally, we summarize our results 
in Sec.~\ref{sec5}. The Appendices provide details
of the NLO corrections to the perturbative cross section
near threshold.

\section{Perturbative Cross Section and Partonic Threshold \label{sec2}}

We are interested in the hadronic cross section for the production 
of two hadrons $h_{1,2}$,
\beq
H_1(P_1) + H_2(P_2) \to h_1(K_1) +h_2(K_2) +
X \; ,
\eeq
with pair invariant mass
\beq
M^2 \equiv (K_1+K_2)^2 \; .
\eeq
We will consider the cross section differential in
the rapidities $\eta_1,\eta_2$ of the two produced hadrons,
treated as massless, in the c.m.s. of the initial
hadrons, or in their difference and average,
\beqa
\Delta \eta &=& \frac{1}{2}(\eta_1 -\eta_2)  \; , \\[1mm]
\bar{\eta} &=&\frac{1}{2}(\eta_1 +\eta_2) \; .
\eeqa
We will later integrate over regions of rapidity corresponding 
to the relevant experimental coverage.
For sufficiently large $M^2$, the cross section for the process 
can be written in the factorized form
\beeq
M^4\frac{d \sigma^{H_1 H_2\to h_1 h_2 X}}{dM^2 d\Delta \eta  d\bar{\eta} }
&=&\sum_{abcd}
\int_0^1 dx_a dx_b  
dz_c  dz_d  \, f_a^{H_1}
(x_a,\mu_{Fi})f_b^{H_2}(x_b,\mu_{Fi})\,  z_c D_c^{h_1}(z_c,\mu_{Ff})
z_dD_d^{h_2}(z_d,\mu_{Ff}) \nn \\
&& \hspace{10mm}
\times  \, \frac{\hat{m}^4
d \hat{\sigma}^{ab\to cd}}{d\hat{m}^2 d\Delta \eta  d\bar{\eta} } 
\left(
\hat{\tau}, \Delta \eta,\hat{{\eta}},\as(\mu_R),
\frac{\mu_R}{\hat{m}}, \frac{\mu_{Fi}}{\hat{m}}, \frac{\mu_{Ff}}{\hat{m}} 
\right) \; ,
\label{taufac} 
\eeeq
where $\hat{{\eta}}$ is the average rapidity in the partonic
c.m.s., which is related to $\bar\eta$ by
\beq
\hat{\eta}=\bar\eta-\frac{1}{2} \ln 
\left(\frac{x_a}{x_b}\right)  \; . 
\label{baretadef}
\eeq
The quantity $\Delta\eta$ is a difference of rapidities and hence 
boost invariant. It is important to note that the 
rapidities of the hadrons with light-like momenta $K_1$ and $K_2$ are the 
same as those of their light-like parent partons. The average and 
relative rapidities for the hadrons and their parent partons
are also therefore the same, a feature that we will use below.
Furthermore, in Eq.~(\ref{taufac})
the $f^{H_{1,2}}_{a,b}$ are the parton distribution functions
for partons $a,b$ in hadrons $H_{1,2}$ and $D_{c,d}^{h_{1,2}}$ the 
fragmentation functions for partons $c,d$ fragmenting into the 
observed hadrons $h_{1,2}$. The distribution functions are 
evaluated at the initial-state and final-state factorization scales  
$\mu_{Fi}$ and $\mu_{Ff}$, respectively. $\mu_R$ denotes the 
renormalization scale. 
The $d \hat{\sigma}^{ab\to cd}/d\hat{\tau} d\bar\eta d\Delta \eta$ 
are the partonic differential cross sections for the contributing 
partonic processes $ab\to cdX'$, where $X'$ denotes some additional 
unobserved partonic final state. The partonic momenta are given in terms
of the hadronic ones by $p_a=x_a P_1$, $p_b=x_b P_2$, 
$p_c=K_1/z_c$, $p_d=K_2/z_d$. We introduce a set of variables, some
of which have been used in Eq.\ (\ref{taufac}):
\beqa
S&=&(P_1+P_2)^2 \; , \\[1mm]
\tau &\equiv& \frac{M^2}{S}  \; , \\[1mm]
\hat{s} &\equiv& \left( x_a P_1 + x_b P_2 \right)^2 = 
x_a x_b S\; , \\[1mm]
\hat{m}^2 &\equiv& \left( \frac{K_1}{z_c} + \frac{K_2}{z_d}\right)^2
= \frac{M^2}{z_cz_d} \; , \\[1mm]
\hat{\tau}&\equiv&\frac{\hat{m}^2}{\hat{s}} =
\frac{M^2}{x_ax_bz_cz_dS} = \frac{\tau}{x_ax_bz_cz_d} \; .
\eeqa
At the level of partonic scattering in the factorized cross section, 
Eq.\ (\ref{taufac}), the other relevant variables are the partonic 
c.m.s. energy $\sqrt{\hat s}$, and the invariant mass 
$\hat m$ of the pair of partons that fragment into the observed di-hadron 
pair. We have written Eq.\ (\ref{taufac}) in such a way that the
term in square brackets is a dimensionless function. 
Hence, it can be chosen to be a function of the dimensionless 
ratio $\hat{m}^2/\hat{s}=\hat{\tau}$ and the ratio of $\hat{m}$ 
to the factorization and renormalization scales, as well as the 
rapidities and the strong coupling. In the following, we will take all 
factorization scales to be equal to the renormalization scale for 
simplicity, that is, $\mu_R=\mu_{Fi}=\mu_{Ff}\equiv\mu$. We then write 
\beqa
\frac{\hat{m}^4
d \hat{\sigma}^{ab\to cd}}{d\hat{m}^2 d\Delta \eta  d\bar{\eta} } 
\left(
\hat{\tau}, \Delta \eta,\hat{\eta},\as(\mu),
\frac{\mu}{\hat{m}} \right)
\equiv 
\omega_{ab\to cd} \left(\th, \Delta \eta, 
\hat{\eta}, \as(\mu), \frac{\mu}{\hat{m}} \right)\; .
\label{omdef}
\eeqa
The variable $\hat{\tau}$ is of special interest for threshold 
resummation, because
it is a measure of the phase space available for 
radiation at short distances. The limit $\hat{\tau} \to 1$ corresponds 
to the partonic threshold, where the partonic hard scattering uses all
available energy to produce the pair. This is kinematically similar to 
the Drell-Yan process, if one thinks of the hadron pair replaced by
a lepton pair. The presence of fragmentation of course complicates the 
analysis somewhat, because only a fraction $z_cz_d$ of $\hat{m}^2$
is used for the invariant mass of the observed hadron pair. In the 
following it will in fact be convenient to also use the variable 
\beqa
\tau' \equiv \frac{\hat{m}^2}{S}= \frac{M^2}{z_cz_dS}\; ,
\label{tauprime}
\eeqa
which is the ratio of the partonic $\hat m^2$ to the overall 
c.m.s. invariant $S$ and hence may be viewed as the
``$\tau$-variable'' at the level of produced partons when fragmentation
has not yet been taken into account. This variable is close in 
spirit to the variable $\tau=Q^2/S$ in Drell-Yan. 

The partonic cross sections can be computed in QCD perturbation theory,
where they are expanded as
\beq
\omega_{ab\to cd} = \left( \frac{\alpha_s}{\pi}\right)^2 \left[ 
\omega_{ab\to cd}^{\mathrm{LO}} + \frac{\alpha_s}{\pi}
\omega_{ab\to cd}^{\mathrm{NLO}} + \ldots \right] \; .
\eeq
Here we have separated the overall power of ${\cal{O}}(\as^2)$, 
which arises because the leading order (LO) partonic hard-scattering 
processes are the ordinary $2\to 2$ QCD scatterings.
At LO, one has $\hat\tau=1$, and also the
two partons are produced back-to-back in the partonic c.m.s., 
so that $\hat{{\eta}}=0$. One can therefore write the LO term as 
\beq
\omega_{ab\to cd}^{\mathrm{LO}} \left(\th, \Delta \eta, 
\hat{{\eta}}\right)=
\delta\left( 1-\hat{\tau}\right)\, \delta \left(\hat{{\eta}}
\right)\,
\omega_{ab\to cd}^{(0)}(\Delta\eta) \; ,
\label{loomega}
\eeq
where $\omega_{ab\to cd}^{(0)}$ is a function of $\Delta \eta$
only. The second delta-function implies that 
$\bar\eta=\frac{1}{2} \ln(x_a/x_b)$.
At next-to-leading order (NLO), or overall ${\cal{O}}(\as^3)$, 
one can have $\hat{\tau}\neq 1$ and $\hat{{\eta}}\neq 0$. 
Near partonic threshold, $\hat\tau\to 1$, however, the kinematics becomes 
``LO like''. The average rapidity of the final-state partons, $c$ and $d$ 
(and therefore of the observed di-hadrons) is determined 
by the ratio $x_a/x_b$, up to corrections that vanish when the 
energy available for soft radiation is squeezed to zero.   
As noted in Ref.\ \cite{Laenen:1992ey}, in this limit the delta function 
that fixes the partonic pair rapidity $\hat{\eta}$ becomes independent
of soft radiation, and may be factored out of the phase space
integral over the latter. This is true at all orders in perturbation
theory. One has:
\beq
\omega_{ab\to cd} \left(\th, \Delta \eta, 
\hat{\eta}, \alpha_s(\mu), \mu/\hat{m} \right)
= \delta \left(\hat{{\eta}} \right)\,
\omega^{{\mathrm{sing}}}_{ab\to cd}\left(\th, \Delta \eta, 
\alpha_s(\mu),\mu/\hat{m} \right)
+\omega_{ab\to cd}^{{\mathrm{reg}}} \left(\th, \Delta \eta, 
\hat{\eta}, \alpha_s(\mu), \mu/\hat{m} \right)
\; , \label{allord}
\eeq
where all singular behavior near threshold is contained in
the functions $\omega^{{\mathrm{sing}}}_{ab\to cd}$. Threshold 
resummation addresses this singular part to all orders in the 
strong coupling. All remaining contributions, which are subleading 
near threshold, are collected in the ``regular'' 
functions $\omega^{{\mathrm{reg}}}_{ab\to cd}$. Specifically,
for the NLO corrections, one finds the following structure:
\beeq
\omega_{ab\to cd}^{\mathrm{NLO}} \left(\th, \Delta \eta, 
\hat{\eta}, \mu/\hat{m} \right)
&=& \delta \left(\hat{{\eta}} \right)\,\left[ 
\omega^{(1,0)}_{ab\to cd}(\Delta \eta, \mu/\hat{m})  \,
\delta (1- \hat{\tau}) \right. \nn \\[2mm]
&&+\left. 
\omega^{(1,1)}_{ab\to cd}(\Delta \eta, \mu/\hat{m}) 
\, \left( \frac{1}{1-\hat{\tau}}  \right)_+  
+ \omega^{(1,2)}_{ab\to cd}(\Delta \eta)
\left( \frac{ \log(1- \hat{\tau}) }{1-\hat{\tau}}  \right)_+  \; \right]
\nn \\[2mm]
&& +\, \omega^{{\mathrm{reg,NLO}}}_{ab\to cd}
(\hat{\tau},\Delta \eta,\hat{\eta}, 
\mu/\hat{m})  \; , \label{NLO}
\eeeq
where the singular part near threshold is represented by the functions
$\omega_{ab\to cd}^{(1,0)}, \omega_{ab\to cd}^{(1,1)}, 
\omega_{ab\to cd}^{(1,2)}$, which are again functions
of only $\Delta\eta$, up to scale dependence. The ``plus''-distributions 
are defined by
\beq
\int_{x_0}^1 f(x)\left( g(x)\right)_+ 
dx\equiv \int_{x_0}^1 \left (f(x) -f(1) \right) \, 
g(x) dx - f(1) \int_0^{x_0} g(x) dx\; .
\eeq
Appendix~A describes the derivation of the coefficients 
$\omega_{ab\to cd}^{(1,0)}, \omega_{ab\to cd}^{(1,1)}, 
\omega_{ab\to cd}^{(1,2)}$ explicitly from a calculation of the 
NLO corrections near threshold. This will serve as a useful check 
on the correctness of the resummed formula, and also to determine 
certain matching coefficients. 

As suggested above, the structure given in Eq.~(\ref{NLO}) is similar
to that found for the Drell-Yan cross section at NLO. A difference
is that in the inclusive
Drell-Yan case one can integrate over all $\Delta \eta$
to obtain a total cross section. This integration is finite because
the LO process in Drell-Yan is the $s$-channel reaction 
$q\bar{q}\to \ell^+\ell^-$. In the case of di-hadrons, the 
LO QCD processes also have $t$ as well as $u$-channel contributions,
which cause the integral over $\Delta\eta$ to diverge when 
the two hadrons are produced back-to-back with large mass,
but each parallel or anti-parallel to the initial beams. As a result,
one will always need to consider only a finite range in $\Delta\eta$.
This is, of course, not a problem as this is anyway also done in 
experiment. It does, however, require a slightly more elaborate 
analysis for threshold resummation, which we review below.

\section{Threshold Resummation for Di-hadron Pairs \label{sec3}}

\subsection{Hard Scales and Transforms}

The resummation of the logarithmic corrections is organized in
Mellin-$N$ moment space~\cite{dyresum}. 
In moment space, the partonic cross sections 
absorb logarithmic corrections associated with the emission of soft 
and collinear gluons to all orders. Employing appropriate moments, 
which we will identify shortly, we will see that the convolutions 
among the different nonperturbative and perturbative regions in the 
hadronic cross section decouple.

In terms of the dimensionless hard-scattering function introduced
in Eq.~(\ref{omdef}) the hadronic cross section in Eq.\ (\ref{taufac})
becomes
\beeq
M^4 \frac{d \sigma^{H_1 H_2\to h_1 h_2 X}}{dM^2 d\Delta \eta  d\bar{\eta} }
&=& \sum_{abcd} 
\int_0^1 dx_a dx_b\, dz_c \, dz_d \,
f_a^{H_1}(x_a)f_b^{H_2}(x_b)\,z_c D_c^{h_1}(z_c)
z_d D_d^{h_2}(z_d)\, \nn \\[1mm]
&& \hspace{10mm} \times \,\omega_{ab\to cd}
\left( \hat{\tau},  \Delta \eta,\hat{{\eta}},
\as(\mu),\frac{\mu}{\hat{m}}
\right) \, ,
\label{taufac1}
\eeeq
where for simplicity we have dropped the scale dependence of the parton
distributions and fragmentation functions.  
At lowest order, when the hard-scattering function $\omega_{ab\to cd}$
is given by Eq.~(\ref{loomega}), the cross section is found to factorize under 
``double'' moments~\cite{Sterman:2000pt,Mukherjee:2006uu}, 
a Mellin moment with respect to $\tau=M^2/S$ and
a Fourier moment in $\bar\eta=\hat{\eta}+\frac{1}{2} \ln(x_a/x_b)$:
\beqa
&&\hspace*{-1cm}\int_{-\infty}^{\infty} 
d\bar\eta \, {\mathrm{e}}^{i \nu \bar{\eta}}\int_0^1 d\tau\,
\tau^{N-1} M^4 \frac{d \sigma^{H_1 H_2\to h_1 h_2 X}}
{dM^2 d\Delta \eta  d\bar{\eta} }\Bigg|_{\mathrm{LO}} \nn \\[3mm]
&&= \sum_{abcd} \tilde{f}_a^{H_1}(N+1+i\nu/2)\tilde{f}_b^{H_2}(N+1-i\nu/2)
\tilde{D}_c^{h_1}(N+2) \tilde{D}_d^{h_2}(N+2)\,  \nn \\[2mm]
&&\hspace*{4mm}\times
\int_{-\infty}^{\infty}d\hat{\eta}\,
{\mathrm{e}}^{i \nu \hat{\eta}}\int_0^1 
d\hat\tau \,\hat\tau^{N-1} \, 
\delta\left( 1-\hat{\tau}\right)\,\delta \left(\hat{{\eta}}\right)\,
\left( \frac{\alpha_s(\mu)}{\pi}\right)^2 \omega_{ab\to cd}^{(0)}
(\Delta\eta)  \; ,
\label{lomom}
\eeqa
where the Mellin moments of the parton distributions or fragmentation
functions are defined in the usual way, for example
\beq
\tilde{f}_a^{H}(N)\equiv\int_0^1  x^{N-1} f_a^H(x) dx \; .
\eeq
We note that instead of a combined Mellin and Fourier transform
one may equivalently use a suitable double-Mellin transform~\cite{cddf}.
The last two integrals in Eq.~(\ref{lomom}) give the combined
Mellin and Fourier moment of the LO partonic cross section. 
Because of the two delta-functions, they are trivial and just yield 
the $N$ and $\nu$ independent result $(\alpha_s/\pi)^2 
\omega_{ab\to cd}^{(0)}(\Delta\eta)$. One might expect that
this generalizes to higher orders, so that the double moments
\beq
\int_{-\infty}^{\infty}d\hat{\eta}\,
{\mathrm{e}}^{i \nu \hat{\eta}}\int_0^1 
d\hat\tau \,\hat\tau^{N-1} \, \omega_{ab\to cd}
\left(\hat{\tau}, \Delta \eta,\hat{\eta},\alpha_s(\mu),
\frac{\mu}{\hat{m}} \right)
\label{doublemoments}
\eeq
would appear times moments of fragmentation functions.
However, this is impeded by the presence of the renormalization/factorization 
scale $\mu$ which must necessarily enter in a ratio with $\hat{m}
=M/\sqrt{z_cz_d}$. As a result of this dependence on $z_c$ and $z_d$,
the moments $\tilde{D}_c^{h_1}(N+2)$, $\tilde{D}_d^{h_2}(N+2)$ of the
fragmentation functions will no longer be generated, and the factorized 
cross section does not separate
into a product under moments.   
Physically, this is a reflection of the mismatch between
the {\it observed} scale, the di-hadron mass $M$, and the
{\it unobserved} threshold scale at the hard scattering, $\hat{m}$.
Threshold logarithms appear when $\hat{s}$ approaches the
latter scale, not the former. This implies that at fixed $M$ there
is actually a range of hard-scattering partonic thresholds, 
extending all the way from $M$ at the lower end to $\sqrt{S}$ at the 
upper. This situation is to be contrasted to the Drell-Yan process 
or to di-jet production at fixed masses, where the underlying hard scale is
defined directly by the observable.

We will deal with the presence of this range of hard scales $\hat{m}$ by
carrying out threshold resummation 
at fixed $\hat m$ as well as at fixed factorization/renormalization scale.   
For this purpose, we rewrite the cross section (\ref{taufac1}) in
a form that isolates the fragmentation functions:
\beq
M^4 \frac{d \sigma^{H_1 H_2\to h_1 h_2 X}}{dM^2 d\Delta \eta  d\bar{\eta} }
= \sum_{cd}\int_0^1 d z_c \, d z_d  \, z_c \,D_c^{h_1}(z_c,\mu) \,z_d\, 
D_d^{h_2}(z_d,\mu) \,
\Omega_{H_1 H_2 \to cd} \left( 
\tau', \Delta \eta, \bar{\eta},  \as(\mu), \frac{\mu}{\hat{m}}  
\right) \, ,
\label{Omegainsigma}
\eeq
where again $\tau'=\hat{m}^2/S=\th x_ax_b$ and 
$\Omega_{H_1 H_2\to cd}$ is given by the convolution of the 
parton distribution functions and $\omega_{ab \to cd}$:
\beq
\Omega_{H_1 H_2\to cd}
\left( \tau', \Delta \eta, \bar{\eta},  \as(\mu), \frac{\mu}{\hat{m}}  
\right) 
=\sum_{ab}\int_0^1 d x_a \,d x_b \,
 f_a^{H_1} \left(x_a, \mu \right)\, f_b^{H_2}\left(x_b, \mu \right)  
\omega_{ab\to cd} \left( \hat{\tau},  \Delta \eta,\hat{{\eta}},
\as(\mu),\frac{\mu}{\hat{m}}
\right) \; , \label{Dconv}
\eeq
with $\hat{\eta}=\bar\eta-\frac{1}{2} \ln(x_a/x_b)$ as before. 
At fixed final-state 
{\it partonic} mass $\hat m$, the function $\Omega_{H_1 H_2\to cd}$ 
now has the desired factorization property under Fourier and
Mellin transforms: 
\beqa
\label{Omemom}
&&\hspace*{-1.8cm}\int_{-\infty}^{\infty} 
d\bar\eta \, {\mathrm{e}}^{i \nu \bar{\eta}}\int_0^1 d\tau'\,
\left(\tau'\right)^{N-1} \Omega_{H_1 H_2\to cd}
\left( \tau', \Delta \eta, \bar{\eta},  \as(\mu), \frac{\mu}{\hat{m}}  
\right) \\[2mm]
&&= \sum_{ab} \tilde{f}_a^{H_1}(N+1+i\nu/2,\mu)
\tilde{f}_b^{H_2}(N+1-i\nu/2,\mu)
\; \tilde{\omega}_{ab\to cd}
\left(N,\nu, \Delta \eta, \as(\mu), \frac{\mu}{\hat{m}} 
\right)  \; , \nn
\eeqa
where
\beq
\tilde{\omega}_{ab\to cd}
\left(N,\nu, \Delta \eta, \as(\mu), \frac{\mu}{\hat{m}} 
\right) \equiv \int_{-\infty}^{\infty}d\hat{\eta}\,
{\mathrm{e}}^{i \nu \hat{\eta}}\int_0^1 
d\hat\tau \,\hat\tau^{N-1} \, \omega_{ab\to cd}
\left(\hat{\tau}, \Delta \eta,\hat{\eta},\alpha_s(\mu),
\frac{\mu}{\hat{m}} \right) \; .
\label{omegamom1}
\eeq
Through Eqs.~(\ref{Omegainsigma})--(\ref{omegamom1})
we have formulated the hadronic cross section in a way that
involves moment-space expressions for the partonic 
hard-scattering functions, which may be resummed. 
Because the final-state fractions $z_i$ equal
unity at partonic threshold, the scale $\hat{m}$ in the short-distance
function may be identified here with the final-state partonic 
invariant mass, up to corrections that are suppressed
by powers of $N$.   For the singular, resummed short-distance
function we therefore do not encounter the problem with the
moments discussed above in connection with Eq.\ (\ref{doublemoments}).

\subsection{Resummation at Next-to-Leading Logarithm \label{sec32}}

As we saw in Eq.~(\ref{allord}), the singular parts of the 
partonic cross sections near threshold enter with 
$\delta(\hat\eta)$. This gives for the corresponding moment-space 
expression
\beq
\tilde{\omega}_{ab\to cd}^{\mathrm{resum}}
\left(N,\Delta \eta, \as(\mu), \frac{\mu}{\hat{m}} 
\right) = \int_0^1 d\hat\tau \,\hat\tau^{N-1} \, 
\omega_{ab\to cd}^{\mathrm{sing}}
\left(\hat{\tau}, \Delta \eta,\alpha_s(\mu),
\frac{\mu}{\hat{m}} \right) \; .
\label{omegamom2}
\eeq
which is a function of $N$ only, but not of the Fourier 
variable $\nu$. Dependence on the Fourier variable $\nu$ then
resides entirely in the parton distributions. It is this function, 
$\tilde{\omega}_{ab\to cd}^{\mathrm{resum}}$, that threshold 
resummation addresses, which is the reason for the use of the
label ``resum'' from now on.

The nature of singularities at partonic threshold is determined by the
available phase space for radiation as $\th\to 1$. Denoting by $k^\mu$
the combined momentum of all radiation, whether from the 
incoming partons $a$ and $b$ or the outgoing partons $c$ and $d$,
one has
\beq
1 - \th = 1 - \frac{(p_c+p_d)^2}{(p_a+p_b)^2} = 
1 - \frac{(p_a+p_b-k)^2}{(p_a+p_b)^2}
\approx \frac{2k^*_0}{\sqrt{s}}\, ,
\label{softenergy}
\eeq
where $k^*_0$ is the energy of the soft radiation in the c.m.s of
the initial partons.

At partonic threshold, the cross section factorizes into ``jet'' functions
associated with the two incoming and outgoing partons, in addition to an
overall soft matrix, traced against the color matrix describing the 
hard scattering~\cite{KS,KOS}. 
Corrections to this factorized structure are suppressed
by powers of $1-\th$. The total cross section is a convolution in 
energy between these functions, which is factorized into a product by 
moments in $\th^N \sim \exp[-N(1-\th)]$, again with corrections 
suppressed by powers of $(1-\th)$, or equivalently, powers of $N$.
This result was demonstrated for jet cross sections 
in \cite{KOS}, and the extension to observed hadrons in 
the final state was discussed in \cite{Sterman:2006hu,Cacciari:2001cw}. 
The resummed expression for the partonic hard-scattering function for the 
process $ab\to cd$ then reads~\cite{KS,BCMN,KOS,KO1}:
\beeq
\tilde{\omega}_{ab\to cd}^{\mathrm{resum}}
\left(N,\Delta \eta, \as(\mu), \frac{\mu}{\hat{m}} 
\right) &=& 
\Delta^{N+1}_a \left(\as(\mu), \frac{\mu}{\hat{m}} \right)
\Delta^{N+1}_b \left(\as(\mu), \frac{\mu}{\hat{m}} \right)\nn \\[2mm]
&&\times\,{\mathrm{Tr}} \left\{ H {\cal{S}}^\dagger_N S  {\cal{S}}_N 
\right\}_{ab \to cd}\left(\Delta \eta, \as(\mu), \frac{\mu}{\hat{m}} 
\right) \nn \\[2mm]
&&\times\, \Delta^{N+2}_c \left(\as(\mu), \frac{\mu}{\hat{m}} \right)
\Delta^{N+2}_d\left(\as(\mu), \frac{\mu}{\hat{m}} \right) \; .
\label{resumm}
\eeeq
We will now discuss 
each of the functions and give their expansions to next-to-leading
logarithmic (NLL) accuracy.

The $\Delta_i^N$ ($i=a,b,c,d$) represent the effects of 
soft-gluon radiation collinear to an initial or final parton. 
Working in the $\overline{\rm{MS}}$ scheme, one 
has~\cite{dyresum,KS,BCMN,KOS,KO1}:
\beeq
\ln \Delta_i^N\left(\as(\mu), \frac{\mu}{\hat{m}} \right)
&=&  \int_0^1 \f{z^{N-1}-1}{1-z} 
\int_{\hat{m}^2}^{(1-z)^2 \hat{m}^2} \f{dq^2}{q^2} A_i(\as(q^2))\nn \\[2mm]
&+&\int_{\mu^2}^{\hat{m}^2}\f{dq^2}{q^2} \left[ -A_i(\as(q^2))\ln\bar{N}-
\frac{1}{2}B_i(\as(q^2)) \right] \; .
\label{Dfct}
\eeeq
Here the functions $A_i$ and $B_i$ are perturbative series in $\as$,
\begin{equation}
A_i(\as)=\frac{\as}{\pi} A_i^{(1)} +  
\left( \frac{\as}{\pi}\right)^2 A_i^{(2)} + \ldots \; ,
\end{equation}
and likewise for $B_i$. To NLL, one needs the coefficients~\cite{KT}:
\beeq 
\label{A12coef} 
&&A_i^{(1)}= C_i
\;,\;\;\;\; A_a^{(2)}=\frac{1}{2} \; C_i  \left[ 
C_A \left( \frac{67}{18} - \frac{\pi^2}{6} \right)  
- \frac{5}{9} N_f \right] \; , \nn \\[2mm] 
&&B_q^{(1)}=-\frac{3}{2} C_F \;,\;\;\;\; B_g^{(1)}=-2\pi \b0 \; ,
\eeeq 
where $N_f$ is the number of flavors, and 
\beeq 
&&C_q=C_F=\frac{N_c^2-1}{2N_c}=\frac{4}{3}  
\;, \;\;\;C_g=C_A=N_c=3 \; , \nn \\[2mm]
&&b_0=\frac{11 C_A - 2 N_f}{12\pi} \; .
\eeeq
The factors $\Delta_i^N$ generate leading threshold enhancements,
due to soft-collinear radiation. We note that our expression for the 
$\Delta_i^N$ differs by the $N$-independent
term proportional to $B_i^{(1)}$ from
that often used in studies of threshold resummation (see, for example, 
Refs.~\cite{BCMN,top}). As was shown in~\cite{KS,KOS,KO1}, this
term is part of the resummed expression and exponentiates. In fact,
the second term on the right-hand-side of Eq.~(\ref{Dfct}) contains 
the large-$N$ part of the moments of the diagonal quark and gluon 
splitting functions, matching the full leading power $\mu_F$-dependence 
of the parton distributions and fragmentation functions in 
Eqs.\ (\ref{Omegainsigma}) and  (\ref{Omemom}). We shall return to this point 
below.

Each of the functions $H_{ab\to cd}$, ${\cal S}_{N,ab\to cd}$, 
$S_{ab\to cd}$ in Eq.~(\ref{resumm})
is a matrix in a space of color exchange operators
\cite{KS,KOS}, and the trace is taken in this 
space. Note that this part is the only one in the resummed
expression Eq.~(\ref{resumm}) that carries dependence on $\Delta \eta$.
The $H_{ab\to cd}$ are the hard-scattering functions.
They are perturbative and have the expansion
\beq
H_{ab\to cd}\left(\Delta\eta,\as(\mu), \frac{\mu}{\hat{m}} \right)=
H_{ab\to cd}^{(0)}\left(\Delta\eta\right)+
\frac{\alpha_s(\mu)}{\pi}
H_{ab\to cd}^{(1)}\left(\Delta\eta,\frac{\mu}{\hat{m}} \right)+
{\cal O}(\alpha_s^2) \; .
\eeq
The LO ({\it i.e.} ${\cal O}(\as^2)$)
parts $H_{ab\to cd}^{(0)}$ are known~\cite{KS,KOS,KO1}, but the 
first-order corrections have not been derived yet. We shall
return to this point shortly. The $S_{ab\to cd}$ are soft functions.
They depend on $N$ only through the argument of the running coupling,
which is set to $\mu/N$ \cite{KS}, and have the expansion
\beq
S_{ab\to cd}\left(\Delta\eta,\as, \frac{\mu}{\hat{m}} \right)=
S_{ab\to cd}^{(0)}+
\frac{\alpha_s}{\pi}
S_{ab\to cd}^{(1)}\left(\Delta\eta,\frac{\mu}{N\hat{m}} \right)+
{\cal O}(\alpha_s^2) \; .
\eeq
The $N$-dependence of the soft function enters the resummed
cross section at the level of next-to-next-to-leading logarithms.
The LO terms $S_{ab\to cd}^{(0)}$ may also be found in~\cite{KS,KOS,KO1}. 
They are independent of $\Delta\eta$.

The resummation of wide-angle soft gluons is contained in
the ${\cal{S}}_{ab\to cd}$, which are exponentials and given
in terms of soft anomalous dimensions, $\Gamma_{ab\to cd}$:
\beeq
\label{GammaSoft}
{\cal S}_{N,ab\to cd} \left(\Delta\eta,\as(\mu), \frac{\mu}{\hat{m}} 
\right) &=& {\cal P}\exp\left[ 
\frac{1}{2} \int^{\hat{m}^2/\bar{N}^2}_{\hat{m}^2} 
\frac{d q^2}{q^2} \Gamma_{ab\to cd} 
\left(\Delta\eta,\as(q^2) \right)\right] \, ,
\eeeq
where ${\cal P}$ denotes path ordering and
where $\bar{N}\equiv N {\mathrm{e}}^{\gamma_E}$ with 
$\gamma_E$ is the Euler constant.
The soft anomalous dimension matrices start at ${\cal O}(\alpha_s)$,
\beq
\Gamma_{ab\to cd}\left(\Delta\eta,\as \right)
=\frac{\as}{\pi}\,\Gamma^{(1)}_{ab\to cd}\left(\Delta\eta\right)+
{\cal O}(\as^2) \; .
\eeq
Their first-order terms are presented in~\cite{KS,KOS,KO1,msj}. 

Note that the Born cross sections are recovered by computing 
${\mathrm{Tr}}\{H^{(0)} S^{(0)}\}_{ab\to cd}$, which is 
proportional to the function $\omega_{ab\to cd}^{(0)}(\Delta\eta)$
introduced in Eq.~(\ref{loomega}). It is instructive to consider 
the expansion of the trace part in Eq.~(\ref{resumm}) to first order
in $\alpha_s$. One finds~\cite{Kidonakis:2001nj}:
\beeq
{\mathrm{Tr}} \left\{ H {\cal{S}}_N^\dagger
S  {\cal{S}}_N \right\}_{ab \to cd} &=&
{\mathrm{Tr}}\{H^{(0)} S^{(0)}\}_{ab\to cd}+
\frac{\alpha_s}{\pi}\,{\mathrm{Tr}}\left\{ -\left[ 
H^{(0)} (\Gamma^{(1)})^\dagger   S^{(0)}+ H^{(0)} S^{(0)}\Gamma^{(1)}  
\right]\ln \bar{N} \right. \nn \\[2mm]
&&\left.+ H^{(1)} S^{(0)}+ H^{(0)} S^{(1)}\right\}_{ab\to cd}
+{\cal O}(\alpha_s^2)\; .
\eeeq
When combined with the first-order expansion of the factors
$\Delta_i^N$ in Eq.~(\ref{resumm}), one obtains
\beeq
\tilde{\omega}_{ab\to cd}^{\mathrm{resum}}
\left(N,\Delta \eta, \as(\mu), \frac{\mu}{\hat{m}} 
\right) &=& {\mathrm{Tr}}\{H^{(0)} S^{(0)}\}_{ab\to cd}
\left( 1 + \frac{\alpha_s}{\pi}\sum_{i=a,b,c,d}A_i^{(1)}
\left[ \ln^2\bar{N} + \ln\bar{N}\ln(\mu^2/\hat{m}^2) \right]
\right)\nn \\[2mm]
&&
+
\frac{\alpha_s}{\pi}\,{\mathrm{Tr}}\left\{ -\left[ 
H^{(0)} (\Gamma^{(1)})^\dagger  S^{(0)}+ H^{(0)} S^{(0)}\Gamma^{(1)} 
\right]\ln \bar{N} \right. \nn \\[2mm]
&&\left.+ H^{(1)} S^{(0)}+ H^{(0)} S^{(1)}\right\}_{ab\to cd}
+{\cal O}(\alpha_s^2)\; .
\label{nloexp}
\eeeq 
This expression can be compared to the results of the explicit
NLO calculation near threshold given in Appendix~A. This provides
a cross-check on the terms that are logarithmic in $N$, that is,
singular at threshold. From comparison to the part proportional
to $\delta(1-\th)$
in the NLO expression, one will be able to read off the
combination $(H^{(1)} S^{(0)}+ H^{(0)} S^{(1)})$ in 
Eq.~(\ref{nloexp}). This is, of course, not sufficient to
determine the full first-order matrices $H^{(1)}$ and $S^{(1)}$,
which would be needed to fully evaluate the trace part in 
in Eq.~(\ref{resumm}) to NLL. To derive  $H^{(1)}$ and $S^{(1)}$,
one would need to perform the NLO calculation near threshold
in terms of a color decomposition~\cite{color}, which is beyond the scope
of this work. Instead, we use here
an approximation that has been made in previous studies (see, for
example, Ref.~\cite{ddfwv}), 
\beeq
{\mathrm{Tr}} \left\{ H {\cal{S}}_N^\dagger
S  {\cal{S}}_N \right\}_{ab \to cd} &\approx&
\left(1+\frac{\alpha_s}{\pi}\,C^{(1)}_{ab \to cd}\right)\,
{\mathrm{Tr}} \left\{ H^{(0)} {\cal{S}}_N^\dagger 
S^{(0)} {\cal{S}}_N \right\}_{ab \to cd} \; ,
\eeeq
where
\beq
C^{(1)}_{ab \to cd}\left( \Delta\eta,\mu/\hat{m}\right)
\equiv \frac{{\mathrm{Tr}} \left\{ 
H^{(1)} S^{(0)}+ H^{(0)} S^{(1)}\right\}_{ab\to cd}}{{\mathrm{Tr}} \left\{ 
H^{(0)} S^{(0)}\right\}_{ab\to cd}}
\label{Ccoeff}
\eeq
are referred to as ``$C$-coefficients''. The coefficients
we obtain for the various partonic channels are given in 
Appendix~B. The approximation we have made becomes exact if only
one color configuration contributes
or if all eigenvalues of the soft anomalous dimension matrix are equal. 
By construction, it is also correct to first order in $\alpha_s$. 

We now turn to the explicit NLL expansions of the ingredients 
in the resummed partonic cross section. For the function
$\Delta_i^N$ in Eq.~(\ref{Dfct}) one finds:
\beq\label{DfctNLL}
\ln \Delta_i^N\left(\as(\mu), \frac{\mu}{\hat{m}} \right)
=  h_i^{(1)} (\lambda)\,\ln \bar{N} + h_i^{(2)} 
\left(\lambda,\as(\mu), \frac{\mu}{\hat{m}} \right) +
\ln{\cal E}_i\left(\lambda,\as(\mu), \frac{\mu}{\hat{m}} \right)\; ,
\eeq
where 
$\lambda = b_0 \as (\mu) \ln \bar{N}$ 
and the functions $h_i^{(1)},h_i^{(2)},\ln ({\cal E}_i)$ are given by
\beeq
h_i^{(1)}(\lambda) &=&\f{A_i^{(1)}}{ 2 \pi \b0 \lambda}
\left( 2 \lambda + \ln ( 1- 2 \lambda) \right) \; ,\nn \\[1mm]
h_i^{(2)}\left(\lambda,\as(\mu), \frac{\mu}{\hat{m}} \right)
& =&\frac{2 \lambda + \ln ( 1- 2 \lambda)}{2\pi b_0}\,
\left( \frac{A_i^{(1)}b_1}{b_0^2} - \f{A_i^{(2)}}{\pi\b0}-
A_i^{(1)}\ln \f{\mu^2}{\hat{m}^2}\right) 
\nn \\[1mm]
&&+
\frac{A_i^{(1)}b_1}{4\pi \b0^3}\ln^2(1-2\lambda)
+\frac{B_i^{(1)}}{2\pi\b0}\ln(1-2\lambda)\; ,
\nn \\[1mm]
\ln{\cal E}_i\left(\lambda,\as(\mu), \frac{\mu}{\hat{m}} \right)&=&
\frac{1}{\pi\b0}\left( -A_i^{(1)}\ln\bar{N}-\frac{1}{2}B_i^{(1)}\right)
\left[ \ln(1-2\lambda) -\b0 \alpha_s(\mu) \ln \f{\mu^2}{\hat{m}^2}
\right] \; .\label{EE}
\eeeq
We note that we have written Eq.~(\ref{DfctNLL}) in a ``non-standard''
form that is actually somewhat more complex than necessary. For example,
one can immediately see that the terms proportional to $B_i^{(1)}
\ln(1-2\lambda)$ cancel between the functions $h_i^{(2)}$ and 
$\ln ({\cal E}_i)$, as they must because they were not present in the 
$\Delta_i^N$ in Eq.~(\ref{Dfct}) in the first place. 
The term proportional to
$\ln(\mu^2/\hat{m}^2)$ in $\ln ({\cal E}_i)$
is the expansion of the second
term in Eq.~(\ref{Dfct}). Its contribution involving $B_i^{(1)}$
does not carry logarithmic dependence on $N$ and would normally be part of 
the ``$C$-coefficients'' discussed above. The term proportional to 
$\ln(1-2 \lambda)$ in $\ln ({\cal E}_i)$ has been separated off the first term
in Eq.~(\ref{Dfct}). Our motivation to use this form of Eq.~(\ref{DfctNLL})
is that the piece termed $\ln({\cal E}_i)$ may be viewed as resulting 
from a large-$N$ leading-order evolution of the corresponding parton 
distribution or fragmentation function between scales
$\hat{m}/\bar{N}$ and the factorization scale $\mu_F$ (we remind the
reader that we have set the factorization and renormalization scales
equal and denoted them by $\mu$). As mentioned earlier, the 
factors $( -2A_i^{(1)}\ln\bar{N}-
B_i^{(1)})$ correspond to the moments of the flavor-diagonal splitting 
functions, $P^N_{ii}$, while the term in square brackets is a LO 
approximation to 
\beq
\b0\int_{\mu_F}^{\hat{m}^2/\bar{N}^2}\frac{dq^2}{q^2}\,\alpha_s(q^2)\; .
\eeq
Therefore, it is natural to identify \cite{KSV} 
\beq
{\cal E}_i\left(\lambda,\as(\mu), \frac{\mu}{\hat{m}} \right)
\,
\tilde{f}_i^H(N,\mu)\,\leftrightarrow\,\tilde{f}_i^H(N,\hat{m}/\bar{N})
\; ,
\label{identify}
\eeq
that is, the exponential related to ${\cal E}_i$ evolves the 
parton distributions from the factorization scale to the scale
$\hat{m}/\bar{N}$, and likewise for the fragmentation functions.
At the level of diagonal evolution, it makes of course no difference 
if $\ln({\cal E}_i)$ is used to evolve the parton distributions or 
if it is just added to the function $h_i^{(2)}$. However, as was
discussed in~\cite{KSV,Catani}, one can actually promote the diagonal
evolution expressed by ${\cal E}_i$ to the full singlet case
by replacing the term $( -2A_i^{(1)}\ln\bar{N}-B_i^{(1)})$
by the full matrix of the moments of the LO singlet splitting functions,
$P_{ij}^{(1),N}$, so that ${\cal E}$ itself becomes a matrix. Using this
matrix in Eq.~(\ref{DfctNLL}) instead of the diagonal ${\cal E}_i$,
one takes into account terms that are suppressed as $1/N$ or higher.
In particular, one resums terms of the form $\as^k\ln^{2k-1}\bar{N}/N$ 
to all orders in $\as$~\cite{Catani}. We will mostly stick to 
the ordinary resummation based on a diagonal evolution operator
${\cal E}_i$ in this paper. However, as we shall show later in one
example, the subleading terms taken into account by implementing the
non-diagonal evolution in the parton distributions and fragmentation
functions can actually be quite relevant in 
kinematic regimes where one is further away from threshold.
Here we will only take the LO part of evolution into account,
extension to NLO is possible and has been discussed in~\cite{KSV}.

For a complete NLL resummation one also needs the expansion of the 
integral in Eq.~(\ref{GammaSoft}), which leads to
\beeq
\label{GammaSoft1}
\ln {\cal S}_{N,ab\to cd} \left(\Delta\eta,\as(\mu), \frac{\mu}{\hat{m}} 
\right)&=& \frac{\ln(1 - 2 \lambda)}{2\pi b_0} \;
\Gamma^{(1)}_{ab\to cd}\left(\Delta\eta\right) \; .
\eeeq
As in~\cite{top}, we perform the exponentiation of the matrix
on the right-hand-side numerically, by iterating the exponential
series to an adequately large order.

\subsection{Inverse of the Mellin and Fourier Transform and 
Matching\\
 Procedure}

As we have discussed in detail, the resummation is achieved in Mellin
moment space. In order to obtain a resummed cross section in 
$\tau$ space, one needs an inverse Mellin transform, accompanied
by an inverse Fourier transform that reconstructs the dependence
on $\bar\eta$. The Mellin inverse 
requires a prescription for dealing with the singularity
in the perturbative strong coupling constant in 
Eqs.~(\ref{Dfct}),(\ref{GammaSoft}) or in the 
NLL expansions, Eqs.~(\ref{DfctNLL}),(\ref{EE}). We will use
the {\em Minimal Prescription} developed in Ref.~\cite{Catani:1996yz},
which relies on use of the NLL expanded forms 
Eqs.~(\ref{DfctNLL}),(\ref{EE}), and on choosing
a Mellin contour in complex-$N$ space that lies to the {\it left}
of the poles at $\lambda=1/2$ and $\lambda=1$ in the Mellin integrand.
From Eqs.~(\ref{Omemom}) and~(\ref{omegamom1}), we find 
\beqa \label{Omegainv}
&&\hspace{-1.5cm}\Omega_{H_1 H_2\to cd}^{\mathrm{resum}}
\left( \tau', \Delta \eta, \bar{\eta},  \as(\mu), \frac{\mu}{\hat{m}}  
\right) =\frac{1}{2\pi}\int_{-\infty}^{\infty} 
d\nu \, {\mathrm{e}}^{-i \nu \bar{\eta}}
\int_{C_{MP}-i\infty}^{C_{MP}+i\infty} \frac{dN}{2\pi i}\,
\left(\tau'\right)^{-N} \nn \\[2mm]
&&\times \sum_{ab} 
\tilde{f}_a^{H_1}(N+1+i\nu/2,\mu)\tilde{f}_b^{H_2}(N+1-i\nu/2,\mu)
\; \tilde{\omega}_{ab\to cd}^{\mathrm{resum}}
\left(N,\nu, \Delta \eta, \as(\mu), \frac{\mu}{\hat{m}} 
\right)  \; , 
\eeqa
where the Mellin contour is chosen so that
$b_0\as(\mu_R^2)\ln C_{MP}<1/2$, but all other poles
in the integrand are as usual to the left of the contour. The
result defined by the minimal prescription has the property that 
its perturbative expansion is an asymptotic series that 
has no factorial divergence and therefore
no ``built-in'' power-like ambiguities~\cite{Catani:1996yz}. 
Power corrections may
then be added as phenomenologically required.
For most of our discussion below, the resummed short-distance function 
$\tilde\omega_{ab\to cd}^{\rm resum}$ is specified 
directly by Eqs.\ (\ref{DfctNLL})
and (\ref{EE}). When we refer to ``full singlet evolution'', however,
we make the identification in Eq.\ (\ref{identify}), 
and evolve the parton distributions and fragmentation functions 
to scale $\hat{m}/\bar{N}$.
In this case the exponential in $\tilde\omega_{ab\to cd}^{\rm resum}$
is found from the $h_i^{(1)}$ and $h_i^{(2)}$ terms only
in Eq.\ (\ref{DfctNLL}).

We note that the parton distribution functions in moment space
fall off with an inverse power of the Mellin moment, typically
as $1/N^4$ or faster. This helps very significantly to 
make the inverse Mellin integral in Eq.~(\ref{Omegainv})
numerically stable. In particular, the resulting
functions $\Omega_{H_1 H_2\to cd}^{\rm resum}$ are very well-behaved 
at high $\tau'$. This would be very different if one were 
to invert just the resummed partonic cross sections $\tilde{\omega}_{
ab\to cd}^{\mathrm{resum}}$ and attempt to convolute
the result with the parton distributions. The good behavior
of the $\Omega_{H_1 H_2\to cd}^{\rm resum}$ 
makes it straightforward numerically
to insert them into Eq.~(\ref{Omegainsigma}), where they are convoluted 
with the fragmentation functions 
in terms of momentum fractions $z$ at fixed rapidities.
At this stage, it is straightforward to impose cuts in the transverse
momenta and rapidities of the observed particles. 
This gives the final hadronic
cross section $M^4 d \sigma^{H_1 H_2\to h_1 h_2 X}/dM^2 d\Delta \eta  
d\bar{\eta}$. We note that because of the presence of the Landau pole
and the definition of the Mellin contour in the minimal prescription,
the inverted $\Omega_{H_1 H_2\to cd}^{\rm resum}$ has support at $\tau'>1$,
where it is however decreasing exponentially with $\tau'$. The 
numerical contribution from this region is very small (less than 
1\%) for all of the kinematics relevant for phenomenology. 

When performing the resummation, one of course wants to make full
use of the available fixed-order cross section, which in our case
is NLO (${\cal O}(\as^3)$). Therefore, a matching to this cross 
section is appropriate, which may be achieved by expanding the resummed 
cross section to ${\cal O}(\as^3)$, subtracting the expanded result
from the resummed one, and adding the full NLO cross section.
Schematically:
\beq
\label{hadnres}
d \sigma^{{\mathrm{match}}} = \left( d \sigma^{{\mathrm{resum}}}-
d \sigma^{{\mathrm{resum}}}\Big|_{{\cal O}(\alpha_s^3)}\right)+
d\sigma^{\mathrm{NLO}} \; .
\eeq
In this way, NLO is taken into account in full, and the soft-gluon 
contributions beyond NLO are resummed to NLL. Any double-counting
of perturbative orders is avoided.

\section{Phenomenological Results \label{sec4}}

We now compare our resummed calculations to experimental
di-hadron production data given as functions of the pair mass, $M$. 
These are available from the fixed-target experiments NA24~\cite{na24} 
($pp$ scattering at beam energy $E_p=300$~GeV), 
E711~\cite{e711} (protons with $E_p=800$~GeV on Beryllium), 
and E706~\cite{e706} ($pp$ and $pBe$ with $E_p=500$ and 
$800$~GeV), as well as from the ISR $pp$ collider experiment CCOR~\cite{ccor}
which produced data at $\sqrt{S}=44.8$ and $62.4$~GeV.
The data sets refer to a $\pi^0\pi^0X$ final state, with the exception 
of E711, which measured the final states $h^+ h^+X$, $h^- h^-X$, 
$h^+ h^-X$ with $h$ summed over all possible hadron species.
When presenting our results for this data set, we will
follow~\cite{Owens:2001rr} to consider for simplicity only the 
{\it summed} charged-hadron combination 
$(h^++h^-)(h^++h^-)X$. For this combination also the
information on the fragmentation functions is more reliable
than for individual charge states.

In each of the experimental data sets, kinematic cuts have been 
applied. These are variously on the individual hadron transverse 
momenta $p_{T,i}$ or rapidities $\eta_i$, or on variables that
are defined from both hadrons, $\cos\theta^*$, $Y$, $p_T^{\mathrm{pair}}$. 
Here $\cos\theta^*$ is the mean of the cosines of the angles between the 
observed hadron directions and the closest beam directions,
in a frame where the produced hadrons have equal and opposite 
longitudinal momenta, $p_{T,1}\sinh\eta_1=-p_{T,2}\sinh
\eta_2$~\cite{na24,e711,e706,ccor,Owens:2001rr}. 
This system approximately coincides with the partonic c.m.s. In terms
of the observed transverse momenta and rapidity difference one has: 
\begin{equation}
\cos\theta^*=\frac{1}{2}
\left( \frac{p_{T,1}}{p_{T,2}+p_{T,1}\cosh(2 \Delta\eta)} +
\frac{p_{T,2}}{p_{T,1}+p_{T,2}\cosh(2 \Delta\eta)} \right) 
\sinh(2\Delta\eta) \; . \label{ctdef}
\end{equation}
Furthermore, $Y$ is the rapidity of the pion pair, 
\begin{equation}
Y=\frac{1}{2}\ln\left(\frac{\kappa^0+\kappa^3}{\kappa^0-\kappa^3}\right)
=\bar{\eta} - \frac{1}{2} \ln\left( \frac{p_{T,1}\,{\mathrm{e}}^{-\Delta\eta} +
p_{T,2}\,{\mathrm{e}}^{\Delta\eta}}{p_{T,1}\,{\mathrm{e}}^{\Delta\eta} +
p_{T,2}\,{\mathrm{e}}^{-\Delta\eta}} \right) \; ,
\end{equation}
where $\kappa=K_1+K_2$ is the pair's four-momentum and where
the second equality in terms of $\Delta \eta$, $\bar\eta$ and the 
hadron transverse momenta $p_{T,i}$ holds for LO kinematics as 
appropriate in the threshold regime.
Finally, $p_T^{\mathrm{pair}}$ is the transverse momentum
of the pion pair, 
\begin{equation}
p_T^{\mathrm{pair}}=|{\bf{p}}_{T,1}+{\bf{p}}_{T,2}|=|p_{T,1}-p_{T,2}| \; ,
\end{equation}
where again the second equality holds to LO.
Thanks to our way of organizing the threshold resummed cross 
section, inclusion of cuts on any of these variables is
straightforward.

In all our calculations, we use the CTEQ6M5 set of parton distribution
functions~\cite{cteq6}, along with its associated value of the
strong coupling constant. We furthermore for the most part use the 
``de Florian-Sassot-Stratmann'' (DSS) fragmentation functions~\cite{DSS},
but will also include comparisons to the results obtained
for the most recent ``Albino-Kniehl-Kramer'' (AKK) set~\cite{akk}.
We note that one might argue that the use of NLO parton distribution functions 
and fragmentation functions is not completely justified for obtaining resummed
predictions, given that large-$N$ resummation effects are typically not 
included in their extraction mostly from
deeply-inelastic scattering (DIS) and $e^+e^-$ annihilation data, 
respectively. As was shown in Ref.~\cite{Shimizu:2005fp} for the case
of the Drell-Yan process, resummation effects in the parton distribution
functions extracted from DIS appear to have a very modest impact, except when 
high momentum fractions and/or relatively low scales 
are probed, which is not the case for
the data sets we are considering here. We expect the same to hold 
for the fragmentation functions. In fact, some large-$N$ resummation 
effects have been included in the AKK analysis~\cite{akk}, and 
comparisons to the results obtained for this set will therefore
be interesting.

We choose for our
calculations the renormalization and factorization scales to be equal,
and we give them the values $M$ and $2M$, in order to investigate
the scale dependence of the results. 
One expects that a natural scale choice would
be offered by the hard scale in the partonic scattering, which 
is ${\cal O}(\hat{m})$. Because of the relation $M=\hat{m}\sqrt{z_cz_d}$,
the scale $M$ is actually significantly lower than $\hat{m}$, 
typically by a factor 2. Our scale choices of $M$ and $2M$
therefore roughly correspond to scales $\hat{m}/2$ and $\hat{m}$,
and we refrain from using a scale lower than $\mu=M$ since this
would correspond to a rather low scale at the partonic hard
scattering.

%%%%%%%%%%%%%%%%%%%%%%%%%%%%%%%%%%%%%%%%%%%%%%%%%%%%%%%%%%%%%%%%%%%%%%
\begin{figure}[p]
\begin{center}
\vspace*{0.6cm}
\epsfig{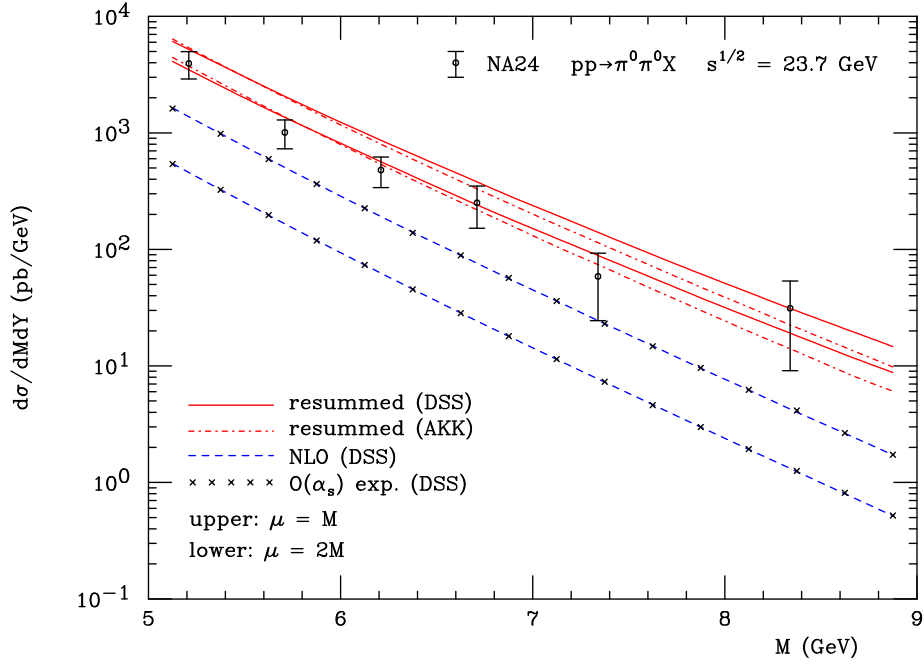}
\end{center}
\vspace*{-.5cm}
\caption{{\it Comparison of the NLO (dashed) and resummed (solid (DSS)
and dash-dotted (AKK)) 
calculations to the NA24 data~\cite{na24}, for two different choices
of the renormalization and factorization scales, $\mu=M$ (upper
lines) and $\mu=2M$ (lower lines). The crosses display the NLO
${\cal O}(\alpha_s)$ expansion of the resummed cross section.
\label{fig1} }}
\vspace*{0.cm}
\end{figure}
%%%%%%%%%%%%%%%%%%%%%%%%%%%%%%%%%%%%%%%%%%%%%%%%%%%%%%%%%%%%%%%%%%%%%%
%%%%%%%%%%%%%%%%%%%%%%%%%%%%%%%%%%%%%%%%%%%%%%%%%%%%%%%%%%%%%%%%%%%%%%
\begin{figure}
\begin{center}
\vspace*{0.6cm}
\epsfig{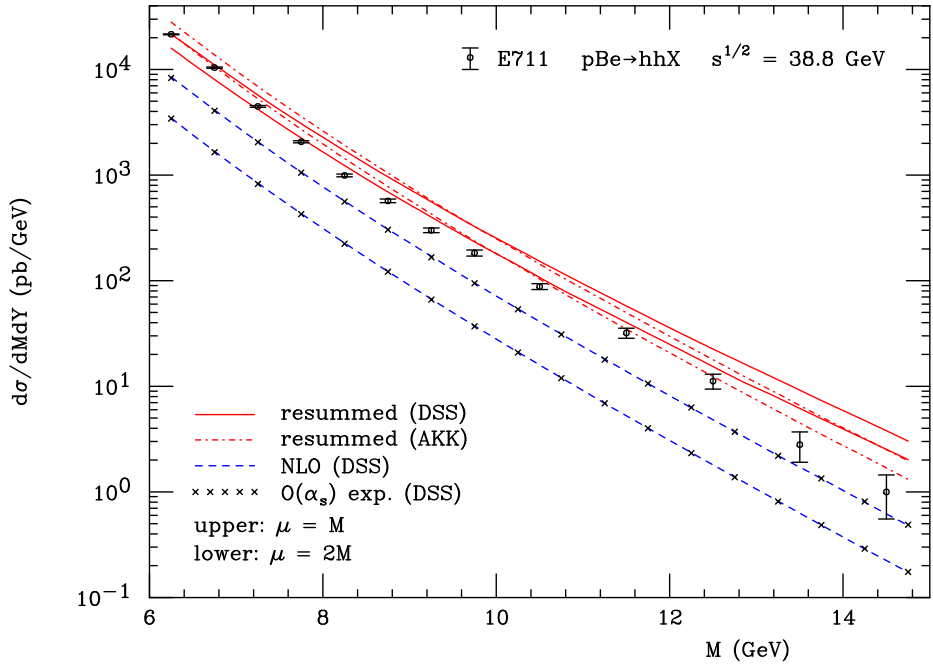}
\end{center}
\vspace*{-.5cm}
\caption{{\it Same as Fig.~\ref{fig1}, but for charged-hadron production
for $pp$ scattering at $\sqrt{S}=38.8$~GeV and with cuts appropriate 
for comparison to E711. The data are from~\cite{e711}.
\label{fig2} }}
\vspace*{0.cm}
\end{figure}
%%%%%%%%%%%%%%%%%%%%%%%%%%%%%%%%%%%%%%%%%%%%%%%%%%%%%%%%%%%%%%%%%%%%%%

Figure~\ref{fig1} shows the comparison to the NA24~\cite{na24} data
for $pp\to \pi^0\pi^0X$ at $\sqrt{S}=23.7$~GeV. 
The cuts employed by NA24 are $|\cos\theta^*|<0.4$, average over
$|Y|<0.35$, and $p_T^{\mathrm{pair}}<1$~GeV. 
We start by comparing the full NLO cross section to the first-order
expansion of the resummed expression, that is, the last two terms in
Eq.~(\ref{hadnres}). This will help to gauge to what extent the
soft-gluon terms constitute the dominant part of the cross section,
so that their resummation is reliable. It turns out that the two
terms agree to a remarkable degree. The dashed lines
in Fig.~\ref{fig1} show the NLO cross section for scales $2M$ (lower)
and $M$ (upper), while the crosses give the NLO expansion of the 
resummed cross section. Their difference actually never 
exceeds $1\%$ for the kinematics relevant for NA24. 
The solid and dash-dotted lines in the figure present the full, 
and matched, resummed results for the DSS and AKK fragmentation 
sets, respectively, including ``$C$-coefficients'' implemented as 
described in Sec.~\ref{sec32} (see Eq.~(\ref{Ccoeff})). 
One can see that resummation leads to a very significant
enhancement of the theoretical prediction. A very good description of 
the NA24 data~\cite{na24} is obtained for both sets, much better 
than for the NLO calculation which falls short of 
the data unless rather low renormalization and factorization scales 
are used. Also the scale dependence of the calculated cross section 
is much reduced by resummation. We note that the resummed result
for the AKK set shows a somewhat steeper $M$-dependence than that
for the DSS set and lies lower at high $M$. This may in part be due 
to the fact that large-$N$ resummation effects were included
in the AKK analysis of the $e^+e^-$ annihilation data, resulting
probably in fragmentation functions that have an overall steeper
$z$-dependence. That said, given the still relatively large 
uncertainties of fragmentation functions overall, 
we also note that the different behavior
of the AKK and DSS results might be just due to differing assumptions 
made in the respective analyses. 

We next turn to the cross section for charged-hadron production,
$pBe\to h^\pm h^\pm X$, measured by E711~\cite{e711} at 
$\sqrt{S}=38.8$~GeV. We recall that we sum over the charges of the 
produced hadrons. The cuts applied by E711 were $p_{T,i}>2$~GeV,
and average over $-0.4<|Y|<0.2$. The cut on the individual hadron 
transverse momenta is, in fact, irrelevant for the values of $M$ 
considered here. Furthermore, as stated in their Fig.~6~\cite{e711} 
for the pair mass distribution we apply $p_T^{\mathrm{pair}}<2$~GeV, 
and $0.1<|\cos\theta^*|<0.25$. Figure~\ref{fig2}
shows the data and our results. As before, the agreement
between NLO and the NLO expansion of the resummed calculation
is excellent. Again, resummation leads to an increase of the
predicted cross section and a reduction of scale dependence. Even
though the resummed results agree with the data much better than
the NLO ones for the scales we have chosen, they tend to lie somewhat
above the data, in particular at the highest values of $M$.
Keeping in mind the results for NA24, one may wonder if this
might be in part related to the fragmentation functions for summed charged 
hadrons, which are probably slightly less well understood than those for 
pions, due to the contributions from the heavier kaons and, in
particular, baryons. The trend for the resummed result to lie a bit 
high is, however, somewhat less pronounced for the AKK set which
again produces results that are a bit steeper than the DSS ones.

\begin{figure}[p]
\begin{center}
\vspace*{0.6cm}
\epsfig{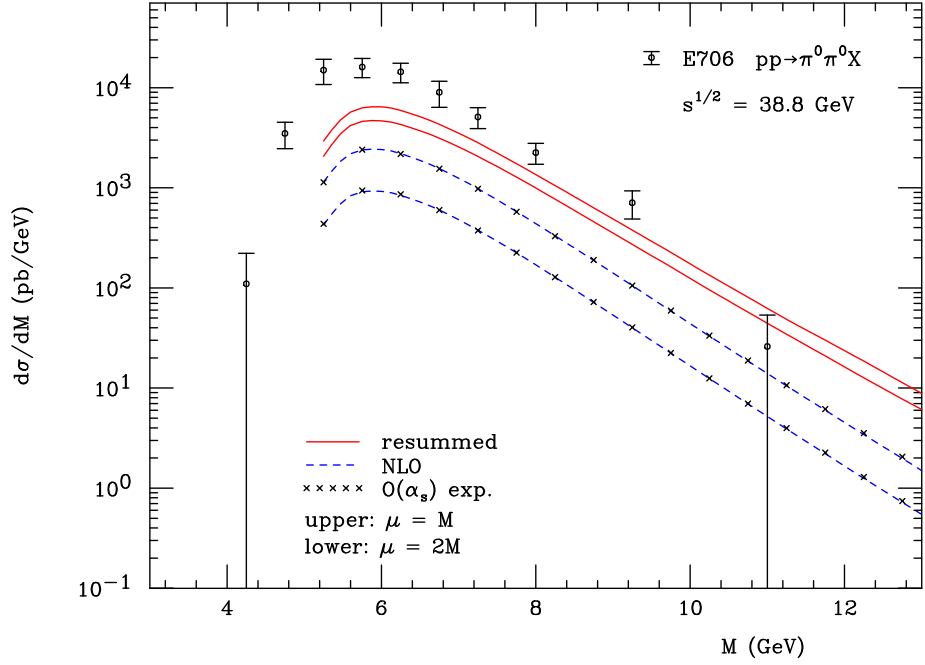}
\end{center}
\vspace*{-.5cm}
\caption{{\it Comparison of the NLO (dashed) and resummed (solid)
calculations (for the DSS fragmentation set)
to the E706 $pp$ 
data at $\sqrt{S}=38.8$~GeV~\cite{e706}, for two different choices
of the renormalization and factorization scales, $\mu=M$ (upper
lines) and $\mu=2M$ (lower lines). The crosses display the NLO
${\cal O}(\alpha_s)$ expansion of the resummed cross section.
\label{fig3} }}
\vspace*{0.cm}
\end{figure}

\begin{figure}
\begin{center}
\vspace*{0.6cm}
\epsfig{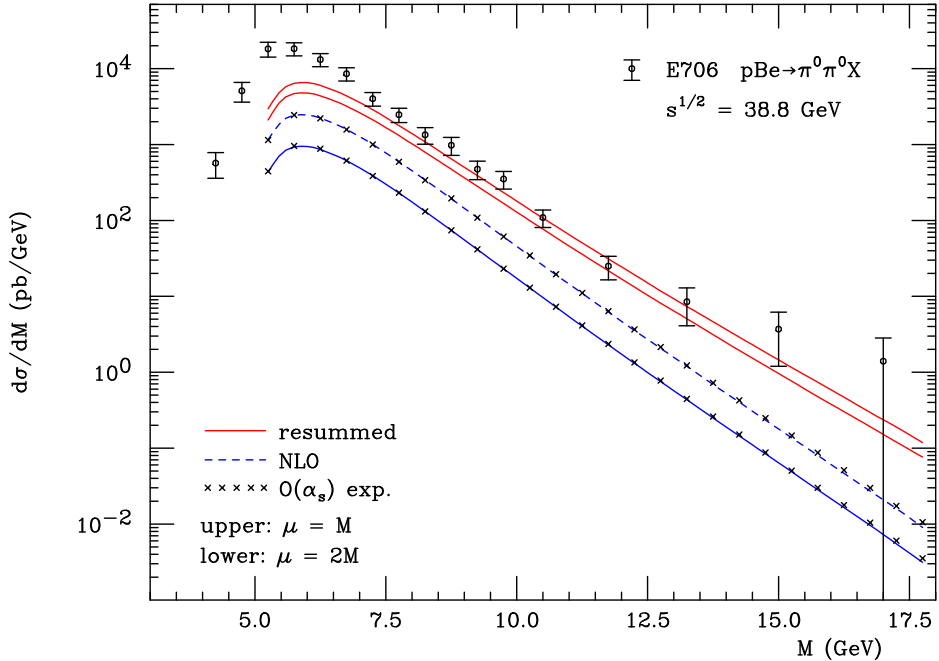}
\end{center}
\vspace*{-.5cm}
\caption{{\it Same as Fig.~\ref{fig3}, but for proton-Beryllium scattering.
\label{fig4} }}
\vspace*{0.cm}
\end{figure}

Figures~\ref{fig3} and~\ref{fig4} show the comparison of our
results (for the DSS set) to the E706 data sets for neutral pion pair
production in $pp$ and $pBe$ scattering at $\sqrt{S}=38.8$~GeV
(800~GeV beam energy),
respectively. We do not take into account any nuclear effects
for the Beryllium nucleus, except for the trivial isospin one.
This has a very minor effect on the cross section, compared
to $pp$. E706 used cuts fairly different from those applied in
the data we have discussed so far. There were no explicit cuts on
$\cos\theta^*$, $p_T^{\mathrm{pair}}$ or $Y$, but instead cuts
$p_{T,i}>p_T^{\mathrm{cut}}=2.5$~GeV and either $-1.05<\eta_i<0.55$ (for the
$\sqrt{S}=38.8$~GeV data) or $-0.8<\eta_i<0.8$ (for the  
$\sqrt{S}=31.6$~GeV data) on the transverse momenta and rapidities
of the individual pions. The cut on transverse momentum, in
particular, has a strong influence at the lower $M$: in a rough
approximation, it leads to a kinematic limit $M\sim 2p_{T,i}>5$~GeV,
so that the cross section has to decrease very rapidly once one
decreases $M$ toward 5~GeV. This behavior is indeed seen in the figures. 

\begin{figure}[p]
\begin{center}
\vspace*{0.6cm}
\epsfig{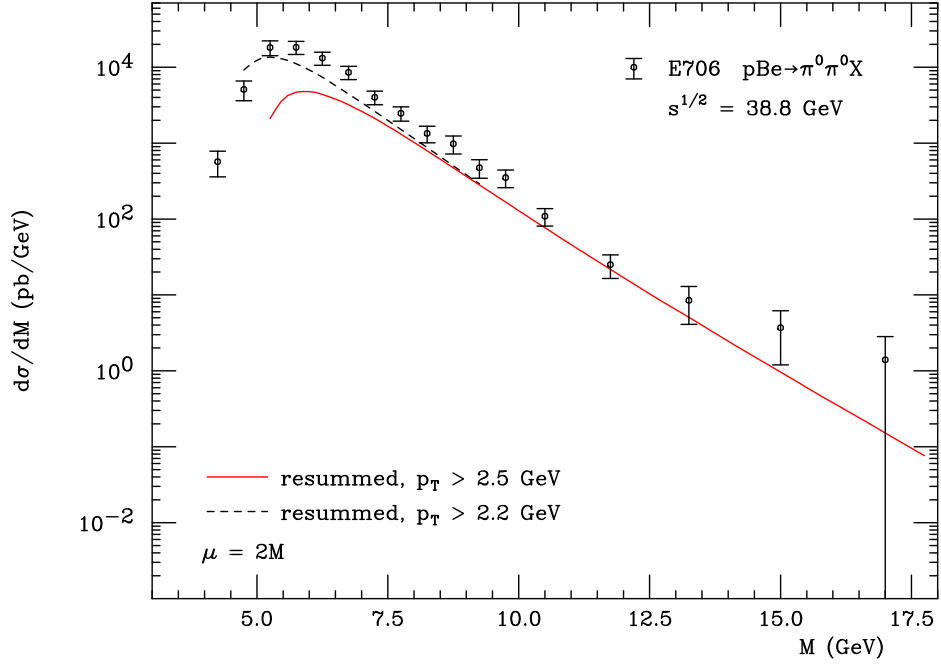}
\end{center}
\vspace*{-.5cm}
\caption{{\it Resummed cross section for scale $\mu=2M$ and 
$p_{T,i}>2.2$~GeV (dashed), compared to the one with 
$p_{T,i}>2.5$~GeV shown previously in Fig.~\ref{fig4} (solid). 
\label{fig5} }}
\vspace*{0.cm}
\end{figure}

As in the previous cases, the NLO expansion of the resummed and
the full NLO cross section agree extremely well, typically to better
than $2\%$. For the two scales we have chosen, the NLO cross sections
fall well short of the data. It was noted in~\cite{Owens:2001rr,Binoth:2002wa} 
that in order for NLO to match the data, very low scales of 
$\mu=0.35M$ have to be chosen. The resummed cross section, on the other
hand, has much reduced scale dependence and describes the data very well
for the more natural scales $M$ and $2M$, except at the lower $M$ where 
the cut $p_T^{\mathrm{cut}}$ on the $p_{T,i}$ becomes relevant. 
One observes that the data extend
to lower $M$ than the theoretical cross section, which basically
cuts off at $M=5$~GeV as discussed above. A new scale becomes
relevant here, the difference $|M-2 p_T^{\mathrm{cut}}|$.
Higher order effects associated with this scale (which are different from
the ones addressed by threshold resummation) and/or non-perturbative
effects such as intrinsic transverse momenta~\cite{e706} probably control
the cross section here. It is also instructive to see
that the cross section is very sensitive to the actual value of the cut 
on the $p_{T,i}$. In Fig.~\ref{fig5} we show the resummed results
for scale $\mu=2M$ for $p_{T,i}>2.5$~GeV (as before) and 
$p_{T,i}>2.2$~GeV. One can see that with the lower cut the 
data are much better described. Experimental resolution effects
might therefore have a significant influence on the comparison 
between data and theory here.

\begin{figure}
\begin{center}
\vspace*{0.6cm}
\epsfig{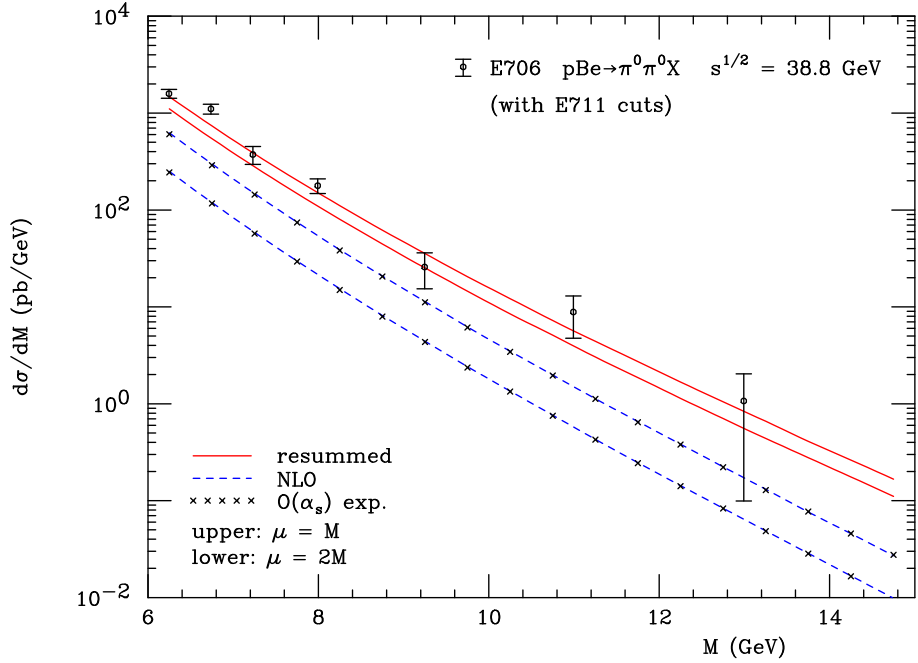}
\end{center}
\vspace*{-.5cm}
\caption{{\it Comparison to E706 data with a different set of cuts,
corresponding to the ones applied by E711. The data with these 
cuts are from~\cite{e706}.  
\label{fig6} }}
\vspace*{0.cm}
\end{figure}

\begin{figure}[p]
\begin{center}
\vspace*{0.6cm}
\epsfig{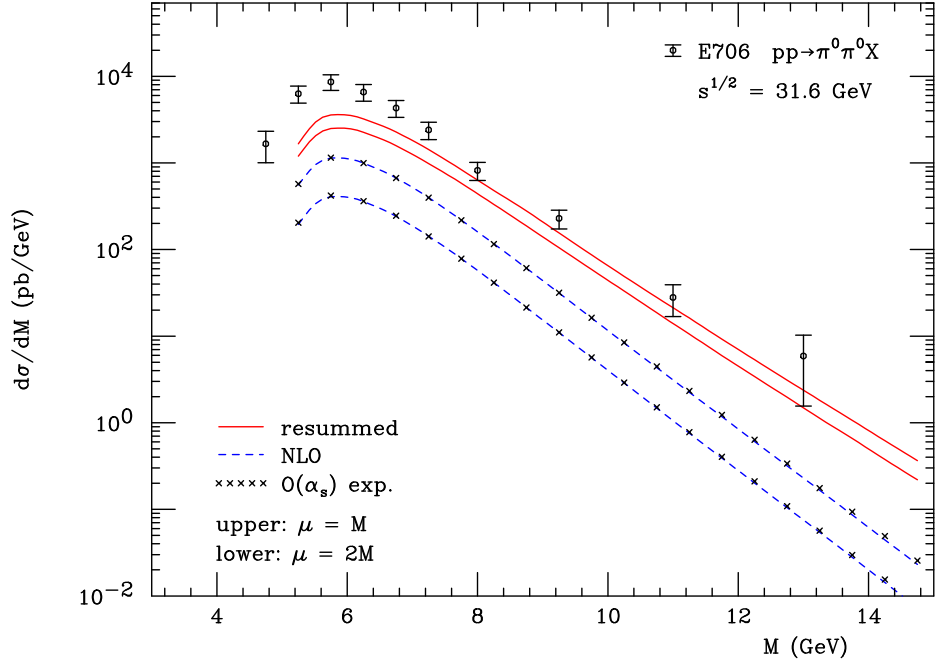}
\end{center}
\vspace*{-.5cm}
\caption{{\it Same as Fig.~\ref{fig3}, but at $\sqrt{S}=31.6$~GeV.
\label{fig7} }}
\vspace*{0.cm}
\end{figure}

\begin{figure}
\begin{center}
\vspace*{0.6cm}
\epsfig{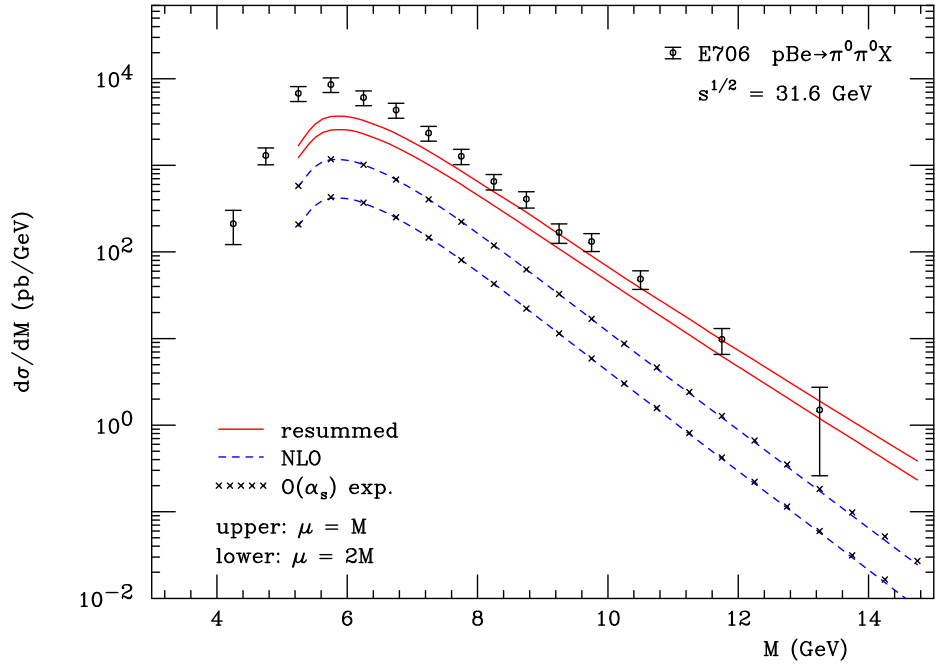}
\end{center}
\vspace*{-.5cm}
\caption{{\it Same as Fig.~\ref{fig4}, but at $\sqrt{S}=31.6$~GeV.
\label{fig8} }}
\vspace*{0.cm}
\end{figure}

In order to check consistency, E706 also presented their $pBe$ data 
set at $\sqrt{S}=38.8$~GeV when the E711 cuts were applied instead of 
the E706 default ones. These data are found in~\cite{e706}. 
Figure~\ref{fig6} shows the comparison for this case. One can see
the same trends as before. Clearly, the description of the data
by the resummed calculation is excellent. For this set of cuts,
the cross section is not forced to turn down by kinematics at the
lower $M$, and theory and data agree well everywhere. Figures~\ref{fig7} 
and~\ref{fig8} show results corresponding to Figs.~\ref{fig3},~\ref{fig4}, 
but for the lower beam energy, 530~GeV, employed by E706 ($\sqrt{S}=
31.6$~GeV). 

%%%%%%%%%%%%%%%%%%%%%%%%%%%%%%%%%%%%%%%%%%%%%%%%%%%%%%%%%%%%%%%%%%%%%%
%%%%%%%%%%%%%%%%%%%%%%%%%%%%%%%%%%%%%%%%%%%%%%%%%%%%%%%%%%%%%%%%%%%%%%
%%%%%%%%%%%%%%%%%%%%%%%%%%%%%%%%%%%%%%%%%%%%%%%%%%%%%%%%%%%%%%%%%%%%%%
%%%%%%%%%%%%%%%%%%%%%%%%%%%%%%%%%%%%%%%%%%%%%%%%%%%%%%%%%%%%%%%%%%%%%%

%%%%%%%%%%%%%%%%%%%%%%%%%%%%%%%%%%%%%%%%%%%%%%%%%%%%%%%%%%%%%%%%%%%%%%
%%%%%%%%%%%%%%%%%%%%%%%%%%%%%%%%%%%%%%%%%%%%%%%%%%%%%%%%%%%%%%%%%%%%%%
%%%%%%%%%%%%%%%%%%%%%%%%%%%%%%%%%%%%%%%%%%%%%%%%%%%%%%%%%%%%%%%%%%%%%%
%%%%%%%%%%%%%%%%%%%%%%%%%%%%%%%%%%%%%%%%%%%%%%%%%%%%%%%%%%%%%%%%%%%%%%

%%%%%%%%%%%%%%%%%%%%%%%%%%%%%%%%%%%%%%%%%%%%%%%%%%%%%%%%%%%%%%%%%%%%%%
%%%%%%%%%%%%%%%%%%%%%%%%%%%%%%%%%%%%%%%%%%%%%%%%%%%%%%%%%%%%%%%%%%%%%%
%%%%%%%%%%%%%%%%%%%%%%%%%%%%%%%%%%%%%%%%%%%%%%%%%%%%%%%%%%%%%%%%%%%%%%
%%%%%%%%%%%%%%%%%%%%%%%%%%%%%%%%%%%%%%%%%%%%%%%%%%%%%%%%%%%%%%%%%%%%%%

\begin{figure}[p]
\begin{center}
\vspace*{0.6cm}
\epsfig{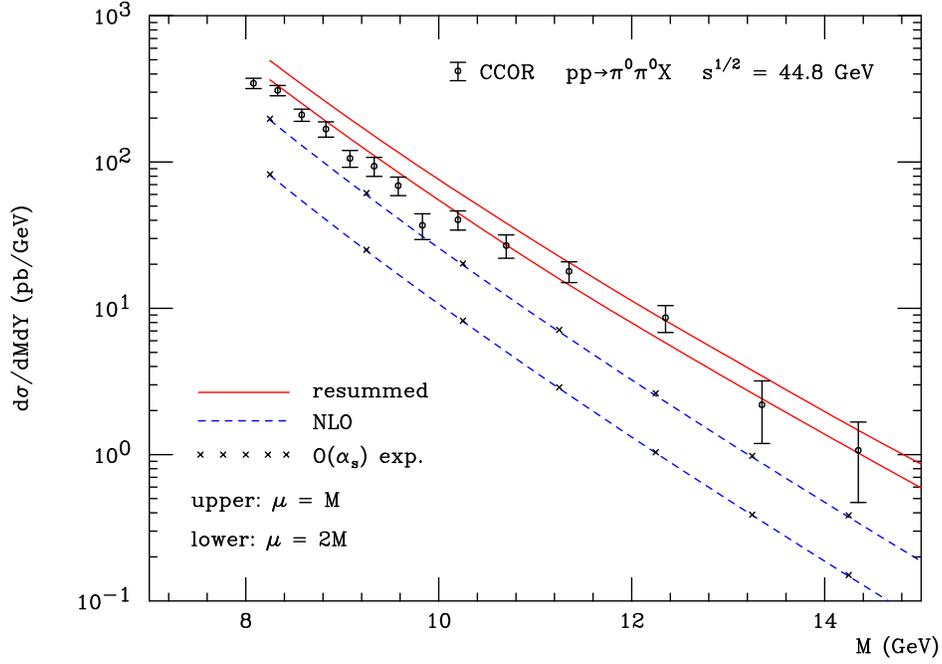}
\end{center}
\vspace*{-.5cm}
\caption{{\it Comparison of the NLO (dashed) and resummed (solid) 
calculations to the CCOR data~\cite{na24} at $\sqrt{S}=44.8$~GeV, 
for two different choices
of the renormalization and factorization scales, $\mu=M$ (upper
lines) and $\mu=2M$ (lower lines). The crosses display the NLO
${\cal O}(\alpha_s)$ expansion of the resummed cross section.
\label{fig9} }}
\vspace*{0.cm}
\end{figure}

\begin{figure}
\begin{center}
\vspace*{0.6cm}
\epsfig{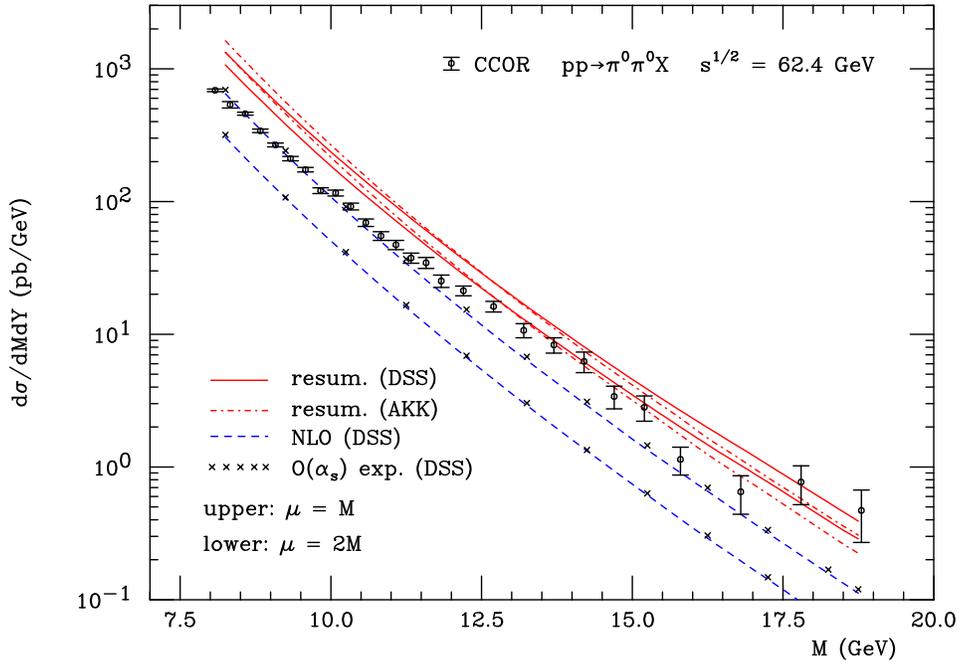}
\end{center}
\vspace*{-.5cm}
\caption{{\it Same as Fig.~\ref{fig9}, but for $\sqrt{S}=62.4$~GeV.
We also show the resummed result obtained for the AKK set
of fragmentation functions. \label{fig10} }}
\vspace*{0.cm}
\end{figure}

We finally turn to the data sets available at the highest energy, which 
are from the CCOR experiment at the ISR~\cite{ccor}. Two data set
are available, at $\sqrt{S}=44.8$~GeV and 62.4~GeV. The cuts employed
by CCOR were identical to those of NA24, $|\cos\theta^*|<0.4$, 
average over $|Y|<0.35$, and $p_T^{\mathrm{pair}}<1$~GeV. 
Figure~\ref{fig9} shows our results at $\sqrt{S}=44.8$~GeV. 
The resummed calculation again shows decreased scale dependence
and describes the data much better than the NLO one. At
the lower values of $M$, it does show a tendency to lie above
the data. Barring any issue with the data (which appear to have a 
certain unexpected ``shoulder'' around $M=10$~GeV or so), this might
indicate that one gets too far from threshold for resummation to
be very precise. On the other hand, the agreement between full NLO
and the NLO expansion of the resummed cross section still remains
very good, as can be seen from the figure. The trend for resummation
to give results higher than the data becomes more pronounced at
the higher energy, $\sqrt{S}=62.4$~GeV, as Fig.~\ref{fig10} shows,
where we have used both the DSS and AKK sets of fragmentation 
functions.
Although not easily seen from the figure, the NLO expansion of
the resummed cross section starts to deviate more from the full NLO
cross section than at the lower energies. At the lower $M$ shown,
it can be higher by up to $7\%$, which is still a relatively minor
deviation, but could be indicative of the reason why the resummed
result is high as well. 

\begin{figure}[p]
\begin{center}
\vspace*{0.6cm}
\epsfig{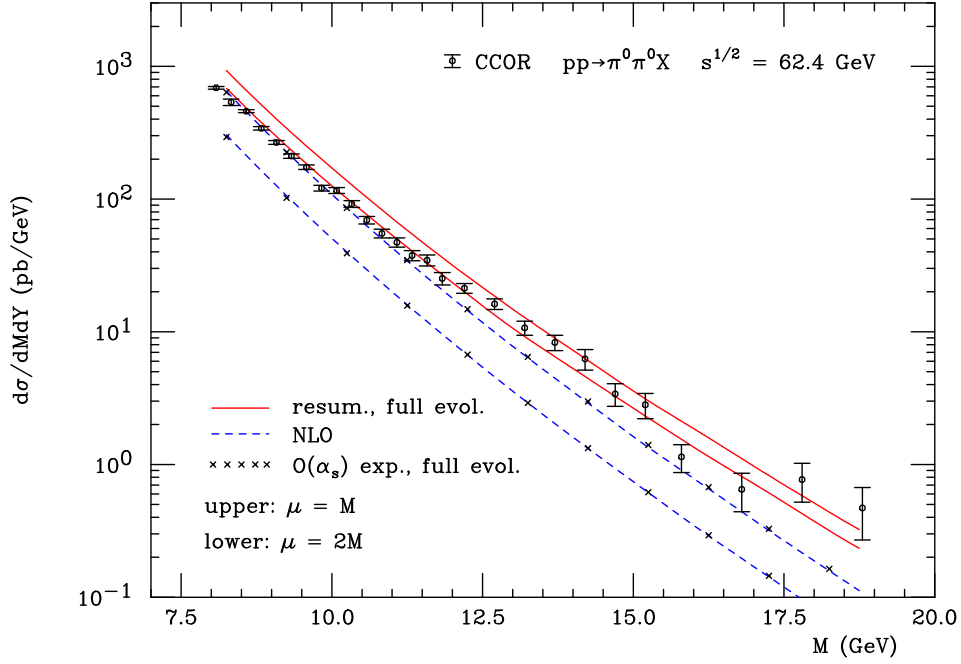}
\end{center}
\vspace*{-.5cm}
\caption{{\it As Fig.~\ref{fig10}, but extending the diagonal
evolution in the resummed formula to included subleading terms
and singlet mixing, as shown in Eq.~(\ref{evolext}). We use
the DSS set of fragmentation functions.
\label{fig11} }}
\vspace*{0.cm}
\end{figure}

\begin{figure}
\begin{center}
\vspace*{0.6cm}
\epsfig{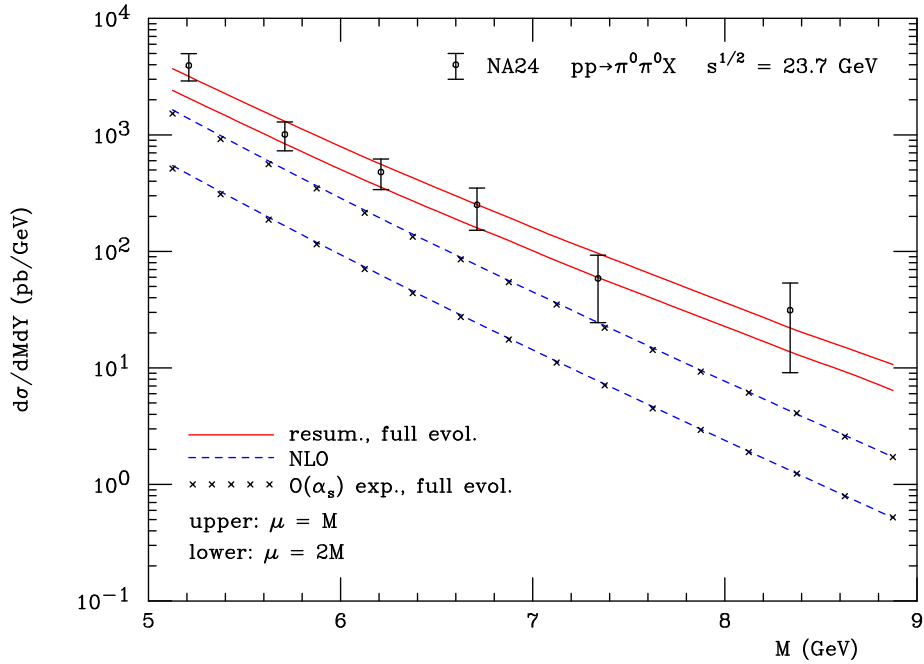}
\end{center}
\vspace*{-.5cm}
\caption{{\it Same as Fig.~\ref{fig11}, but for the case of 
NA24.
\label{fig12} }}
\vspace*{0.cm}
\end{figure}

Clearly, any deviation between the full NLO cross section and 
the NLO expansion of the resummed one is due to terms that
are formally suppressed by an inverse power of the Mellin moment $N$
near threshold. 
It is therefore interesting to explore the
likely effects of such terms. This can be done by promoting 
the LO anomalous dimension in the evolution part in Eq.~(\ref{DfctNLL}) from
its diagonal form to the full one, as described in Sec.~\ref{sec32}:
\begin{equation}
-2A_i^{(1)}\ln\bar{N}-B_i^{(1)}\rightarrow P_{ij}^{(1),N} \; ,
\label{evolext}
\end{equation} 
which includes the subleading terms in $1/N$ and full 
singlet mixing. For simplicity, we perform this modification only
for the lowest order part of evolution, as indicated in Eqs.~(\ref{EE}) 
and (\ref{evolext}). The results obtained in this way are shown
in Fig.~\ref{fig11}. One can see that the resummed result
obtained in this way indeed decreases significantly with
respect to the one in Fig.~\ref{fig10} which was based on
the diagonal evolution only, and is much closer to the data.
At the same time, the agreement between the NLO cross section
and the ${\cal O}(\alpha_s)$ expanded resummed result becomes
as good as what we encountered in the fixed-target case. 
Figure~\ref{fig12} presents the corresponding result for the case
of NA24. Comparison with Fig.~\ref{fig1} shows that the
effect of the subleading terms is much smaller here, as expected
from the fact that one is closer to threshold in the case of
NA24. Nonetheless, the effects lead to a slight further improvement
between the resummed calculation and the data. In particular, they
give the theoretical result a somewhat flatter behavior, which 
follows the trend of the data more closely overall.
While the implementation of subleading terms in this way
will require further study, this appears to be a promising
approach for extending the applicability of threshold resummation
into regimes where one is relatively far away from threshold.

%%%%%%%%%%%%%%%%%%%%%%%%%%%%%%%%%%%%%%%%%%%%%%%%%%%%%%%%%%%%%%%%%%%%%%
%%%%%%%%%%%%%%%%%%%%%%%%%%%%%%%%%%%%%%%%%%%%%%%%%%%%%%%%%%%%%%%%%%%%%%
%%%%%%%%%%%%%%%%%%%%%%%%%%%%%%%%%%%%%%%%%%%%%%%%%%%%%%%%%%%%%%%%%%%%%%
%%%%%%%%%%%%%%%%%%%%%%%%%%%%%%%%%%%%%%%%%%%%%%%%%%%%%%%%%%%%%%%%%%%%%%

%%%%%%%%%%%%%%%%%%%%%%%%%%%%%%%%%%%%%%%%%%%%%%%%%%%%%%%%%%%%%%%%%%%%%%
%%%%%%%%%%%%%%%%%%%%%%%%%%%%%%%%%%%%%%%%%%%%%%%%%%%%%%%%%%%%%%%%%%%%%%
%%%%%%%%%%%%%%%%%%%%%%%%%%%%%%%%%%%%%%%%%%%%%%%%%%%%%%%%%%%%%%%%%%%%%%
%%%%%%%%%%%%%%%%%%%%%%%%%%%%%%%%%%%%%%%%%%%%%%%%%%%%%%%%%%%%%%%%%%%%%%

\begin{figure}[p]
\begin{center}
\vspace*{0.6cm}
\epsfig{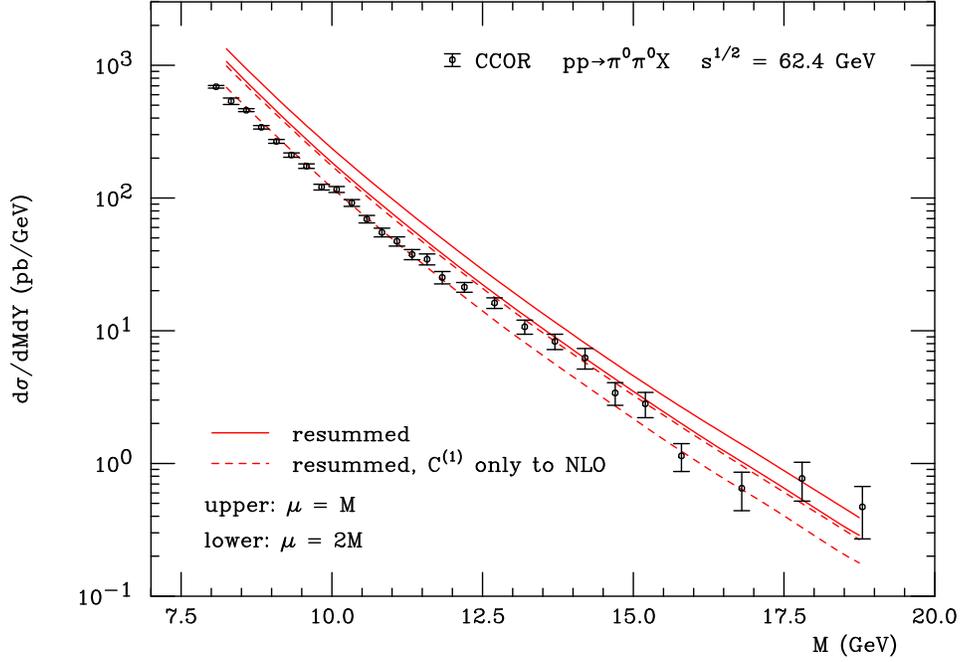}
\end{center}
\vspace*{-.5cm}
\caption{{\it Resummed results for the case of CCOR at $\sqrt{S}=
62.4$~GeV. The solid lines show the results for scales $M$ and $2M$
shown previously in Fig.~\ref{fig10}, while the dashed ones
were obtained by neglecting the contributions by the $C^{(1)}_{ab\to cd}$ 
coefficients beyond NLO.  
\label{fig13} }}
\vspace*{0.cm}
\end{figure}

That said, we remind the reader that already in the part that is leading
near threshold we have made the approximation in Eq.~(\ref{Ccoeff})
for our ``$C$-coefficients''. This, too, will need to be improved
in the future, by taking into account the full color structure of
the hard scattering function beyond LO, as we discussed in 
Sec.~\ref{sec32}. To give a somewhat extreme example of the 
effects generated by the $C$-coefficients, we have re-computed
the resummed cross section for the case of CCOR at $\sqrt{S}=62.4$~GeV,
but leaving out all effects of the the coefficients {\it beyond NLO}.
In other words, we leave out the $C$-coefficients in the
first two terms on the right-hand-side of Eq.~(\ref{hadnres}),
keeping them of course in $d\sigma^{\mathrm{NLO}}$. This is likely
not a good approximation of the beyond-NLO hard coefficients, because
the $C^{(1)}_{ab \to cd}$ have $\pi^2$ terms and logarithms 
in the renormalization scale $\mu$ that are independent of the color 
channel and truly enter in the form given in Eq.~(\ref{Ccoeff}). Some of 
these are in fact even known to 
exponentiate~\cite{KS,KOS,KO1,Kidonakis:2001nj,eric}. In any case,
the result of this exercise is shown in Fig.~\ref{fig13}, where it is 
also compared to our earlier calculation that included the $C$-coefficients
in the way discussed in Sec.~\ref{sec32}. One can see that there
is a sizable numerical difference, and that the scale dependence
of the resummed result without the beyond-NLO $C$-coefficients becomes
significantly worse. 
 
\begin{figure}
\begin{center}
\vspace*{0.6cm}
\epsfig{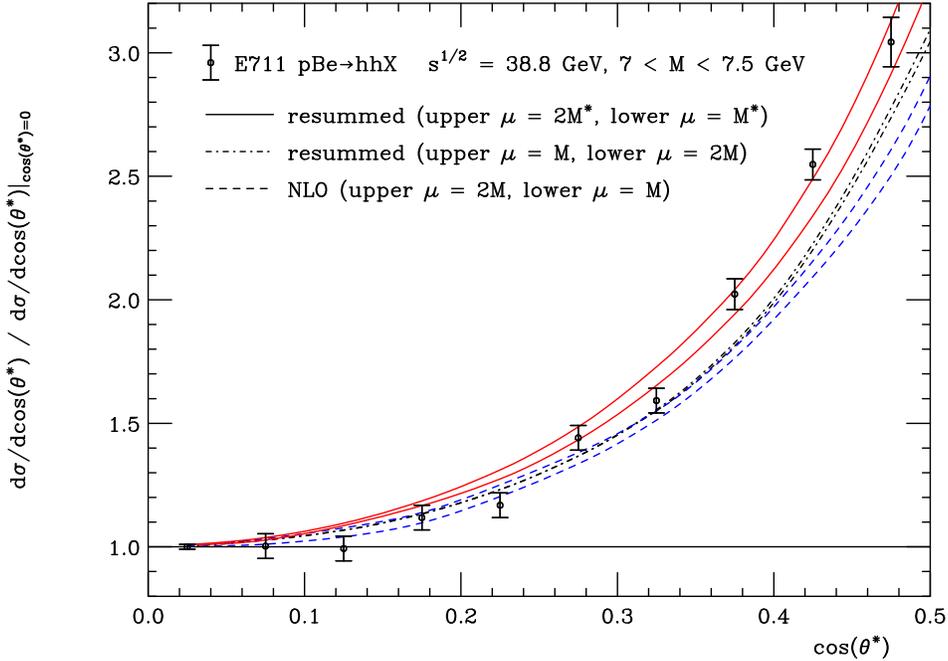}
\end{center}
\vspace*{-.5cm}
\caption{{\it Normalized distribution in $\cos\theta^*$ (see~(\ref{ctdis}))
for the case of charged-hadron production at E711. Dashed is NLO,
while the dot-dashed and solid lines show resummed results. For the 
latter we have also used the scales $\mu=M^*$ and $\mu=2 M^*$, where $M^*{}^2=
M^2 (1-\cos\theta^*)$.   
\label{fig14} }}
\vspace*{0.cm}
\end{figure}

We finally turn to the distribution in $\cos\theta^*$, defined
in Eq.~(\ref{ctdef}), for which most of the experiments mentioned 
above have presented data as well. In fact, the CCOR data~\cite{ccor}
for this observable were instrumental in establishing the QCD 
hard-scattering nature of $pp$ interactions~\cite{mjt}. From the
point of view of threshold resummation, the distribution in 
$\cos\theta^*$ may appear somewhat less interesting than the pair 
mass one, since the threshold logarithms arise in $1-\hat{\tau}=
1-\hat{m}^2/\hat{s}$, regardless of $\cos\theta^*$. In addition, 
the $\cos\theta^*$ distributions are presented as normalized
distributions of the form
\beq
\frac{d\sigma/d\cos\theta^*}{d\sigma/d\cos\theta^*|_{\cos\theta^*=0}} \; ,
\label{ctdis}
\eeq
so that the main enhancement generated by threshold resummation is
expected to cancel. Nonetheless, as we have seen in Sec.~\ref{sec32},
the resummed expressions do contain additional dependence on 
$\Delta\eta$ beyond that present in the Born cross sections, which
will affect the $\cos\theta^*$ distribution at higher orders.
This is visible from the soft part in Eq.~(\ref{GammaSoft1}) and
also from the ``$C$-coefficients'' in Eq.~(\ref{Ccoeff}). 
Rather than going through an exhaustive comparison to all the
available data, we just consider one example that is representative of 
the effects of threshold resummation on the $\cos\theta^*$ distribution.
Figure~\ref{fig14} shows the normalized distribution for the E711 case,
where we have again summed over all charge states of the produced hadrons.
The dashed lines show the NLO result calculated again with the 
code of~\cite{Owens:2001rr}, for scales $\mu=2 M$ and $\mu=M$.
One can see that for these scales the NLO calculation is lower
than the data for higher values of $\cos\theta^*$. 
The dot-dashed lines in Fig.~\ref{fig14} show the resummed results
for scales $\mu=2 M$ and $\mu=M$. These show a steeper rise with
$\cos\theta^*$ and describe the data better than NLO for the scales 
shown. However, they still tend to lie below the data at higher values
of $\cos\theta^*$. As was suggested 
in~\cite{ccor,Owens:2001rr,Binoth:2002wa}, for 
the $\cos\theta^*$ distribution the hard scale in the partonic process 
will itself be a function of $\cos\theta^*$, so that it is more
natural to choose a factorization/renormalization scale that
reflects this feature. We therefore present our resummed results
also for scales $\mu=2 M^*$ and $\mu=M^*$, where $M^*{}^2=
M^2 (1-\cos\theta^*)$ which is proportional to the Mandelstam
variable $\hat{t}$ in the partonic process. One observes that
with these scale choices a very good description of the data is
achieved. We note that in the NLO calculations presented in
Refs.~\cite{Owens:2001rr,Binoth:2002wa} the scale was chosen
proportional to the (average) transverse momenta of the produced
hadrons, which for given $M$ also depend on $\cos\theta^*$. This
resulted in a satisfactory description of the data, when
scales effectively a factor two smaller than our $M^*$ were
used. Overall, the trend for the resummed $\cos\theta^*$ 
distribution to lie higher than NLO and be in better agreement with 
the data is found to be a generic feature that occurs as well for 
the cases of the other experiments. 

%%%%%%%%%%%%%%%%%%%%%%%%%%%%%%%%%%%%%%%%%%%%%%%%%%%%%%%%%%%%%%%%%%%%%%
%%%%%%%%%%%%%%%%%%%%%%%%%%%%%%%%%%%%%%%%%%%%%%%%%%%%%%%%%%%%%%%%%%%%%%
%%%%%%%%%%%%%%%%%%%%%%%%%%%%%%%%%%%%%%%%%%%%%%%%%%%%%%%%%%%%%%%%%%%%%%
%%%%%%%%%%%%%%%%%%%%%%%%%%%%%%%%%%%%%%%%%%%%%%%%%%%%%%%%%%%%%%%%%%%%%%

\section{Conclusions \label{sec5}}

We have investigated the effects of next-to-leading
logarithmic threshold resummation on
the cross section for di-hadron production
in hadronic collisions, $H_1 H_2\to h_1 h_2 X$, for a
range of invariant masses of the produced hadron pair.
We have developed techniques to implement the resummation formalism
at fixed rapidities for the produced hadrons and for all relevant 
experimental cuts. Extensions of these techniques to the level 
of next-to-next-lo-leading logarithms
should be relatively straightforward in light of the close relation
between the one- and two-loop soft anomalous dimension 
matrices~\cite{twoloopad}.

For the fixed target and collider
data studied here, the one-loop
expansions of our resummed expressions approximate the corresponding
exact one-loop cross sections excellently, to the level of a few percent
and often less.
In addition, with scales chosen to match the underlying hard scattering,
the matched resummed cross sections
typically explain the available data 
better than do NLO expressions at similar scales, 
with significantly reduced scale dependence.

An important extension of these methods will
be in the production and fragmentation of heavy quarks
and in jet cross sections, where similar resummation methods are 
applicable. Given the reduction in scale dependence, this could
provide an improved control over Standard Model tests and backgrounds
in new physics searches.

\section*{Acknowledgments} 

We are grateful to M.\ Begel and H.B.\ White 
for very helpful communications on the E706 and E711 data, respectively,
and to S.\ Albino, B.\ J\"{a}ger, A.\ Mitov, M.\ Stratmann and M.\ Tannenbaum 
for useful discussions. We also thank J.F.\ Owens
for providing his NLO code for di-hadron production, and for 
comments. W.V.\ is grateful to the U.S.\ Department of Energy 
(contract number DE-AC02-98CH10886) for
providing the facilities essential for the completion of his work.
This work was supported in part by the National Science Foundation, 
grants PHY-0354776, PHY-0354822 and PHY-0653342.

\section*{Appendix A}
In this appendix we present some details for the calculation 
of the NLO partonic cross-sections near threshold. The virtual
corrections have the $2\to 2$ kinematics of the Born terms
and therefore fully contribute. They are proportional to $\delta
(1-\th)$. The real-emission $2\to 3$ contributions require
more effort. We consider the reaction $a(p_1)+b(p_2)\to 
c(k_1)+d(k_2)+e(k_3)$, where partons $d$ and $e$ fragment
into the observed pair of hadrons and have pair mass $\hat{m}^2$.
It is convenient to work in the c.m.s. of the observed outgoing
hadrons. We can then write the three-body phase 
space in $4-2 \varepsilon$ dimensions as
\beeq \label{ps2}
\Phi_3 &=&\frac{s}{(4\pi)^4\Gamma(1-2 \varepsilon)}
\left( \frac{4 \pi}{s} \right)^{2\varepsilon}
\, \int_0^1 d\hat{\tau}\hat{\tau}^{-\varepsilon}(1-\hat{\tau})^{1-2\varepsilon}
\int_0^\infty d\rho \rho^{-\varepsilon} (1+\rho)^{-2+2 \varepsilon}
\nonumber \\[2mm]
&&\times 
\int_0^{\pi} d\psi \sin^{1-2\varepsilon}\psi \int_0^\pi d\theta\,
\sin^{-2 \varepsilon}\theta \; .
\eeeq
Here we define 
\beq 
\rho=(p_1-k_2)^2/(p_2-k_2)^2=
{\mathrm{e}}^{-2 \Delta\eta} \; .
\eeq
Near threshold, the integration variables are given in terms of the 
Mandelstam variables of the process as follows:
\beeq
&&(p_1+p_2)^2 =\hat{s}  \; , \;\;\;(k_2+k_3)^2=\hat{m}^2=\th\hat{s} 
\; , \nn \\[1mm]
&&(p_1-k_1)^2 = -\frac{\hat{s}(1-\th)}{2}(1-\cos\psi) \; , \;\;\;
(p_2-k_1)^2 = -\frac{\hat{s}(1-\th)}{2}(1+\cos\psi) \; , \nn \\[1mm]
&&(p_1-k_2)^2=-\frac{\hat{s}\rho}{1+\rho}=(p_2-k_3)^2 \; , \;\;\;
(p_2-k_2)^2=-\frac{\hat{s}}{1+\rho}=(p_1-k_3)^2  \; , \nn \\[2mm]
&&(k_1+k_2)^2=\frac{\hat{s}(1-\th)}{2}\left( 1+\sin\psi \cos\theta \,
\frac{2\sqrt{\rho}}{1+\rho} -\cos\psi \,\frac{1-\rho}{1+\rho} \right)\; , 
\nn \\[2mm]
&&(k_1+k_3)^2=\frac{\hat{s}(1-\th)}{2}\left( 1-\sin\psi \cos\theta \,
\frac{2\sqrt{\rho}}{1+\rho} +\cos\psi \,\frac{1-\rho}{1+\rho} \right)\; .
\eeeq
The phase space in Eq.~(\ref{ps2})
is used to integrate the squared $2\to 3$ matrix elements 
$|{\cal{M}}_{ab \to cde}|^2$. For the latter one also assumes
near-threshold kinematics. Since we want the partonic cross section
at fixed $\hat{\tau}$ and $\Delta\eta$, we only need to perform 
the last two integrations in Eq.~(\ref{ps2}).
The basic integral for these is~\cite{WvN}
\beeq
\label{omega}
&&\int_0^\pi d\psi \int_0^\pi d\theta\,
\frac{\sin^{1-2\varepsilon}\psi \sin^{-2 \varepsilon}\theta}
{(1-\cos\psi)^j (1-\cos\psi \cos\chi-\sin\psi \cos\theta 
\sin\chi)^k} \nn \\[2mm]
&&\hspace{1cm}= 2 \pi \frac{\Gamma(1-2 \varepsilon)}{\Gamma(1-\varepsilon)^2}
\,2^{-j-k}\,B(1-\varepsilon-j,1-\varepsilon-k)\,
{}_2 F_1 \left(j,k,1-\varepsilon,\cos^2 \frac{\chi}{2} \right) \; , 
\eeeq
where ${}_2 F_1$ is the Hypergeometric function. After integration
over phase space and addition of the virtual corrections, infrared 
singularities cancel and only collinear singularities remain. These
are removed by mass factorization, which we do in the $\msbar$ scheme.
Notice that since we are close to threshold only the diagonal 
splitting functions $P_{ii}^{(1)}$ contribute in this procedure. 
Combining all contributions, one arrives at the near-threshold structure 
of the partonic cross sections given in Eq.~(\ref{NLO}), for each 
subprocess that is already present at LO. The final step is
to take Mellin moments in $\th$ of the result, as described in
Eq.~(\ref{omegamom1}). This gives for the partonic cross sections
to NLO:
\beeq
\hspace*{-1cm}\tilde{\omega}_{ab\to cd}^{\mathrm{thr,LO+NLO}}
\left(N,\Delta \eta,\alpha_s(\mu), \mu/\hat{m}\right) 
&=& \omega_{ab\to cd}^{(0)}(\Delta\eta) \nn \\[2mm]
&&\hspace*{-6cm}+\frac{\alpha_s(\mu)}{\pi}\left[ 
\omega^{(1,0)}_{ab\to cd}(\Delta \eta, \mu/\hat{m}) \, - \, 
\ln\bar{N}\;\omega^{(1,1)}_{ab\to cd}(\Delta \eta, \mu/\hat{m})
+ \f{1}{2} \left( \ln^2 \bar{N}+\zeta(2) \right)\;
\omega^{(1,2)}_{ab\to cd}(\Delta \eta) \right] \; ,
\eeeq
where terms subleading in $N$ have been neglected. The ``$C$-coefficients''
defined in Eq.~(\ref{Ccoeff}) are obtained from this as 
\beq
C^{(1)}_{ab \to cd}\left( \Delta\eta,\mu/\hat{m}\right)
=\frac{\omega^{(1,0)}_{ab\to cd}(\Delta \eta, \mu/\hat{m}) 
+\frac{1}{2}\zeta(2)\,\omega^{(1,2)}_{ab\to cd}(\Delta \eta)}{
\omega_{ab\to cd}^{(0)}(\Delta\eta)} \; .
\eeq

\section*{Appendix B}
In this section we give the coefficients $C^{(1)}_{ab \to cd}$ for each
subprocess contributing to the production of our di-hadron final state,
resulting from the calculation outlined in Appendix~A. 
In all expressions below, $\mu$ is the renormalization scale.
The dependence on the factorization scale is already included
in the function ${\cal E}_i$ in Eq.~(\ref{EE}).
As before, we define $\rho\equiv{\mathrm{e}}^{-2 \Delta\eta}$.

\noindent
\underline{$qq'\to qq'$:}\\[3mm]
We define:
\begin{equation}
Q_{qq'}\equiv 1+(1+ \rho)^2 \; .
\end{equation}
We then have:
\begin{eqnarray}
C_{qq'\to qq'}^{(1)}\left( \Delta\eta,\mu/\hat{m}\right)&=&
2 \pi b_0 \ln \frac{\mu^2}{\hat{m}^2}+\left(\frac{5}{6 
Q_{qq'}}+\frac{13}{12}\right) \ln^2\rho+
\left(\frac{5}{6}-\frac{1}{3 Q_{qq'}}\right) 
\ln^2(1+\rho)\nonumber \\[2mm]
&&+\left(-\frac{8}{3}+\frac{14+9 \rho}{6 Q_{qq'}}\right) 
\ln\rho+\left(-\frac{4}{3}+\frac{2}{3 Q_{qq'}}\right) \ln
(1+\rho) \ln\rho\nonumber \\[1mm]
&&-\frac{\rho}{3 Q_{qq'}} \ln (1+\rho)+\frac{7 \pi ^2}{6 Q_{qq'}}
+\frac{N_f}{3}  
\ln\frac{\rho}{1+\rho}-\frac{5 
N_f}{9}\nonumber \\[1mm]
&&+\frac{8}{3} 
\text{Li}_2\left(\frac{\rho}{1+\rho}\right)+\frac{3}{2} \ln
(1+\rho)+\frac{47 \pi ^2}{36}+\frac{7}{2} \; .
\end{eqnarray}

\noindent
\underline{$q\bar{q}'\to q\bar{q}'$:}\\[3mm]
We have:
\begin{eqnarray}
C_{q\bar{q}'\to q\bar{q}'}^{(1)}\left( \Delta\eta,\mu/\hat{m}\right)&=&
C_{qq'\to qq'}^{(1)}\left( \Delta\eta,\mu/\hat{m}\right)+\frac{5}{6}\left\{
\left(1-\frac{2}{Q_{qq'}}\right) \left[ (1+\ln\rho)\ln\rho+
\frac{\pi^2}{2} \right]\right.
\nonumber \\[2mm]
&&\left.\hspace*{-1.5cm}-\frac{\rho}{Q_{qq'}} \ln (1+\rho) +
\left(\frac{3}{2}-\frac{1}{Q_{qq'}}\right) \ln(1+\rho) 
\ln\frac{1+\rho}{\rho^2}-2\text{Li}_2\left(\frac{\rho}{1+\rho}
\right)
\right\} \; .
\end{eqnarray}

\noindent
\underline{$qq\to qq$:}\\[3mm]
We define:
\begin{equation}
Q_{qq}\equiv \frac{(1-\rho+\rho^2)(3+5 \rho+3 \rho^2)}{(1+\rho(1+\rho))} \; .
\end{equation}
We then have:
\begin{eqnarray}
C_{qq\to qq}^{(1)}\left( \Delta\eta,\mu/\hat{m}\right)&=&
2 \pi b_0 \ln \frac{\mu^2}{\hat{m}^2}
+\frac{8}{Q_{qq}} \left(1-\rho^2\right) 
\text{Li}_2\left(\frac{\rho}{1+\rho}\right)\nonumber \\[2mm]
&&+\left(\frac{7}{6}-\frac{59 \rho}{48 Q_{qq}}+\frac{5}{4 \
Q_{qq}}-\frac{\rho+4}{16 \left(3 +5 \rho+3\rho^2\right)}
\right) \ln^2\rho\nonumber \\[2mm]
&&-\frac{\left(12 \rho^2+3 \rho-4\right)}{2
Q_{qq}} \ln^2(1+\rho)+\frac{\ln\rho}{12Q_{qq}}
\left(37 \rho-71+\frac{(17-8 \rho)\,Q_{qq}}{3 +5 \rho+3\rho^2}\right) 
\nonumber \\[2mm]
&&+\left(\frac{7}{3}-\frac{7}{4Q_{qq}} \left( 6-5\rho \right)
-\frac{53 \rho-6}{12 \left(3 +5 \rho+3\rho^2\right)}\right) 
\ln(1+\rho) \ln\rho\nonumber \\[2mm]
&&+\left(\frac{3}{2}-\frac{\rho}{4 Q_{qq}}-\frac{\rho}{4 \left(3 
+5 \rho+3\rho^2\right)}\right) \ln(1+\rho)\nonumber \\[2mm]
&&+N_f \left(\frac{2-\rho }{2Q_{qq}} 
+\frac{\rho}{3 \left(3 +5 \rho+3\rho^2\right)}\right) \ln\rho
-\frac{1}{3} N_f \ln (1+\rho)-
\frac{5 N_f}{9}\nonumber \\[2mm]
&&+\frac{7}{2}\left(1+\frac{2}{3}\pi ^2\right)
-\frac{\pi^2}{3 Q_{qq}}\left(4+\frac{41}{16}\rho\right)
-\frac{71 \pi ^2 \rho}{144 \left(3 +5 \rho+3\rho^2\right)} \; .
\end{eqnarray}

\noindent
\underline{$q\bar{q}\to q'\bar{q}'$:}\\[3mm]
We define:
\begin{equation}
Q_{q'\bar{q}'}\equiv 1+\rho^2 \; .
\end{equation}
We then have:
\begin{eqnarray}
C_{q\bar{q}\to q'\bar{q}'}^{(1)}\left( \Delta\eta,\mu/\hat{m}\right)&=&
2 \pi b_0 \ln \frac{\mu^2}{\hat{m}^2}
+\frac{7}{4}\left(1-\frac{2}{3 Q_{q'\bar{q}'}}\right) 
\ln^2\rho-\frac{5}{12}\left(1+\frac{2}{Q_{q'\bar{q}'}}\right) 
\ln^2(1+\rho)\nonumber \\[2mm]
&&+\frac{7 (1+\rho)}{6 Q_{q'\bar{q}'}} \ln\rho
-\frac{7}{6}\left(1-\frac{2}{Q_{q'\bar{q}'}}\right) \ln (1+\rho) \ln\rho
-\frac{1}{3}\left(1+\frac{5+9 \rho}{2Q_{q'\bar{q}'}}\right) 
\ln(1+\rho)\nonumber \\[2mm]
&&-\frac{5 N_f}{9}-\frac{5}{3}
\text{Li}_2\left(\frac{\rho}{1+\rho}\right)+\frac{1}{6} 
\left(21+4 \pi ^2\right) \; .
\end{eqnarray}

\noindent
\underline{$q\bar{q}\to q\bar{q}$:}\\[3mm]
We define:
\begin{eqnarray}
&&Q^{(1)}_{q\bar{q}}\equiv 3+\rho(1+\rho) \; , \nonumber \\
&&Q^{(2)}_{q\bar{q}}\equiv 1+3\rho(1+\rho) \; .
\end{eqnarray}
We then have:
\begin{eqnarray}
C_{q\bar{q}\to q\bar{q}}^{(1)}\left( \Delta\eta,\mu/\hat{m}\right)&=&
2 \pi b_0 \ln \frac{\mu^2}{\hat{m}^2}+
N_f \left(\frac{1}{6}+(1+2\rho)\left(\frac{1}{8 Q^{(1)}_{q\bar{q}}}+
\frac{1}{8 Q^{(2)}_{q\bar{q}}}\right)\right)
\ln\left(\frac{\rho}{1+\rho} \right) \nonumber \\[2mm]
&&\hspace*{-2cm}+\text{Li}_2\left(\frac{\rho}{1+\rho}\right) 
\left(\frac{5+4 \rho}{2 Q^{(1)}_{q\bar{q}}}+\frac{1+4 \rho}{2 
Q^{(2)}_{q\bar{q}}}-\frac{1}{3}\right)+\pi ^2 \left(\frac{5 (9+14 \rho)}{96 
Q^{(1)}_{q\bar{q}}}+\frac{155+282 \rho}{288Q^{(2)}_{q\bar{q}}}+
\frac{43}{36}\right)\nonumber \\[2mm]
&&\hspace*{-2cm}+
\left(\frac{4 \rho-79}{64 Q^{(1)}_{q\bar{q}}}+\frac{61+180 \rho}{576 
Q^{(2)}_{q\bar{q}}}+\frac{65}{36}\right)\ln^2 \rho  
+\left(\frac{13+124 \rho}{64 Q^{(1)}_{q\bar{q}}}+\frac{361+972 \rho}
{576 Q^{(2)}_{q\bar{q}}}+\frac{29}{36}\right)\ln^2(1+\rho) \nonumber \\[2mm]
&&\hspace*{-2cm}+
\left(\frac{7-\rho}{16 Q^{(1)}_{q\bar{q}}}-\frac{35+71 \rho}{48 
Q^{(2)}_{q\bar{q}}}-\frac{11}{12}\right)\ln\rho+
\left(\frac{61-64 \rho}{32 Q^{(1)}_{q\bar{q}}}-\frac{247+576 \rho}{288
Q^{(2)}_{q\bar{q}}}-\frac{22}{9}\right)\ln(1+\rho) \ln \rho\nonumber \\[2mm]
&&\hspace*{-2cm}+
\left(\frac{8+\rho}{16 Q^{(1)}_{q\bar{q}}}+\frac{36+71 \rho}{48 
Q^{(2)}_{q\bar{q}}}+\frac{7}{12}\right)\ln(1+\rho)-\frac{5 N_f}{9}+\frac{7}{2}
\; .
\end{eqnarray}

\noindent
\underline{$q\bar{q}\to gg$:}\\[3mm]
We define:
\begin{equation}
G_{q\bar{q}}\equiv (1+\rho^2)(4-\rho+4 \rho^2) \; .
\end{equation}
We then have:
\begin{eqnarray}
C_{q\bar{q}\to gg}^{(1)}\left( \Delta\eta,\mu/\hat{m}\right)&=&
2 \pi b_0 \ln \frac{\mu^2}{\hat{m}^2}
-\frac{27}{2 G_{q\bar{q}}} \left(1-\rho^4\right)
\text{Li}_2\left(\frac{\rho}{1+\rho}\right)\nonumber \\[2mm]
&&+ \frac{1}{48}\left(1+\frac{2\rho}{G_{q\bar{q}}}\left( 133+13\rho\right)
+\frac{124-311\rho}{4 -\rho+4\rho^2}
\right) \ln^2\rho\nonumber \\[2mm]
&&+\frac{1}{48}\left(69 + \frac{52 \rho^2}{G_{q\bar{q}}}-
\frac{\rho+648}{4 -\rho+4\rho^2}
\right) \ln^2(1+\rho)\nonumber \\[2mm]
&&+\frac{1}{6}\left(-\frac{\rho}{G_{q\bar{q}}}\left(3+
89 \rho\right)+\frac{48+5 \rho}{4 -\rho+4\rho^2}\right)
\ln\rho\nonumber \\[2mm]
&&+\left(\frac{89 \rho^2}{3 G_{q\bar{q}}}-\frac{19 
\rho}{6 \left(4-\rho+4 \rho^2\right)}-2\right) \ln(1+\rho)\nonumber \\[2mm]
&&+\frac{1}{24}\left(-19-\frac{2\rho}{G_{q\bar{q}}}\left(133+13 \rho\right)
+\frac{200+149 \rho}{4 -\rho+4\rho^2}\right) \ln\rho \ln
(1+\rho)\nonumber \\[2mm]
&&-\frac{15}{4G_{q\bar{q}}} \rho (1-\rho)^2+\frac{9 
\pi ^2 (4-\rho)}{16 \left(4-\rho+4 \rho^2\right)}+\frac{191 
\pi ^2}{144}-\frac{14}{3} \; .
\end{eqnarray}

\noindent
\underline{$qg\to qg$:}\\[3mm]
We define:
\begin{eqnarray}
&&Q^{(1)}_{qg}\equiv 2 (1+\rho)+\rho^2 \; , \nonumber \\
&&Q^{(2)}_{qg}\equiv 9 (1+\rho)+4\rho^2 \; .
\end{eqnarray}
We then have:
\begin{eqnarray}
C_{qg\to qg}^{(1)}\left( \Delta\eta,\mu/\hat{m}\right)&=&
2 \pi b_0 \ln \frac{\mu^2}{\hat{m}^2}-\frac{14}{3}+\frac{15 
(1+\rho) (2+\rho)^2}{4 Q^{(1)}_{qg} Q^{(2)}_{qg}} \nonumber \\[2mm]
&&\hspace*{-2cm}+\pi ^2 
\left(\frac{146+13 \rho}{24 Q^{(1)}_{qg}}-\frac{3 (109+13 \rho)}{16 
Q^{(2)}_{qg}}+\frac{241}{144}\right)\nonumber \\[2mm]
&&\hspace*{-2cm}+\left((1+\rho)\left( \frac{13}{12 Q^{(1)}_{qg}}-\frac{15}{16 
Q^{(2)}_{qg}}\right)+\frac{17}{16}\right)\ln^2\rho
+ (1+\rho) \left(\frac{89}{3 Q^{(1)}_{qg}}-\frac{231}{2 Q^{(2)}_{qg}}
\right) \ln\rho \nonumber \\[2mm]
&&\hspace*{-2cm}+\left(\frac{13 \rho-120}{24 Q^{(1)}_{qg}}+
\frac{3 (173+41 \rho)}{16 
Q^{(2)}_{qg}}-\frac{27}{16}\right)\ln^2(1+\rho)\nonumber \\[2mm]
&&\hspace*{-2cm}+\left(-\frac{86+89 \rho}{6 Q^{(1)}_{qg}}
+\frac{3 (43+39 \rho)}{2 Q^{(2)}_{qg}}-2\right)\ln (1+\rho)+
\left(\frac{31}{24}+\frac{27 (\rho-3)}{8 
Q^{(2)}_{qg}}\right) 
\text{Li}_2\left(\frac{\rho}{1+\rho}\right) \nonumber \\[2mm]
&&\hspace*{-2cm}+ \left(\frac{120-13 \rho}{12 Q^{(1)}_{qg}}-
\frac{3 (155+23 \rho)}{8 
Q^{(2)}_{qg}}+\frac{31}{24}\right)\ln\rho\ln (1+\rho) \; .
\end{eqnarray}

\noindent
\underline{$gg\to q\bar{q}$:}\\[3mm]
We have:
\begin{eqnarray}
C_{gg\to q\bar{q}}^{(1)}\left( \Delta\eta,\mu/\hat{m}\right)
&=&C_{q\bar{q}\to gg}^{(1)}\left( \Delta\eta,\mu/\hat{m}\right) \; .
\end{eqnarray}

\noindent
\underline{$gg\to gg$:}\\[3mm]
We define:
\begin{equation}
G_{gg}\equiv 1+\rho(1+\rho) \; .
\end{equation}
We then have:
\begin{eqnarray}
C_{gg\to gg}^{(1)}\left( \Delta\eta,\mu/\hat{m}\right)&=&
2 \pi  b_0 \ln \frac{\mu^2}{\hat{m}^2}+N_f 
\left(\frac{5}{9}+\frac{3 \rho^2 (1+\rho)^2}{8 G_{gg}^3}+\frac{\pi ^2 
\rho \left(1+\rho^2\right) (1+\rho)^2}{16 \
G_{gg}^3}\right)\nonumber \\[2mm]
&&\hspace*{-2cm}-\frac{3 \rho^2 (1+\rho)^2}{8 \
G_{gg}^3}(3+\pi^2)+\frac{3}{4}\left(\frac{(1+\rho)^3}{
G_{gg}^3}+1\right) \ln^2\rho\nonumber \\[2mm]
&&\hspace*{-2cm}+\frac{N_f}{16G_{gg}^2}
(1+\rho) \left(\frac{2}{G_{gg}}(1+\rho)^2
+\rho^2-2 (1+\rho) \right) \ln^2\rho\nonumber \\[2mm]
&&\hspace*{-2cm}+\frac{N_f}{24G_{gg}^2} \left(8\left(1+\rho^2\right)+
5 \rho\right) (1+\rho)^2 
\ln (1+\rho)-\frac{N_f}{24 G_{gg}^2} (1+\rho) \left(5
\rho^2+8(1+\rho)\right) \ln\rho\nonumber \\[2mm]
&&\hspace*{-2cm}+\frac{N_f}{16G_{gg}^2} (1+\rho)^2 
\left(\frac{2}{G_{gg}}(1+\rho)
-2-\rho\right) \ln^2(1+\rho)\nonumber \\[2mm]
&&\hspace*{-2cm}+\frac{N_f}{8G_{gg}^2} (1+\rho) 
\left(2 \rho+1-\frac{1}{G_{gg}}(1+\rho)^2\right) \ln\rho \ln
(1+\rho)\nonumber\\[2mm]
&&\hspace*{-2cm}+\frac{1}{G_{gg}^2}\left(-\frac{11}{2} 
\left(1+\rho^2\right)-\frac{7}{4}\rho\right) (1+\rho)^2 \ln
(1+\rho)+\frac{1}{G_{gg}^2}\left(\frac{7}{4}\rho^2+\frac{11}{2}
(1+\rho)\right) (1+\rho) \ln\rho\nonumber \\[2mm]
&&\hspace*{-2cm}+\frac{3 \pi ^2 (1+\rho)}{4 \
G_{gg}^2}+\left(\frac{3}{2}-\frac{3 (1+\rho)^3}{4 G_{gg}^3}+\frac{3 (2 \rho
+1) \left(1+\rho^2\right)}{4 G_{gg}^2}\right) 
\ln \rho \ln (1+\rho)\nonumber \\[2mm]
&&\hspace*{-2cm}+\left(\frac{3 (1+\rho)^3}{4 \
G_{gg}^3}-\frac{3 (2\rho+1)}{4 G_{gg}^2}-\frac{3 \
(1+\rho)}{2 G_{gg}}-\frac{3}{4}\right) \ln^2(1+\rho)\nonumber\\[2mm]
&&\hspace*{-2cm}-\frac{3}{2G_{gg}}
\left(1-\rho^2\right) \
\text{Li}_2\left(\frac{\rho}{1+\rho}\right)
-\frac{\pi ^2 (1+\rho)}{2 G_{gg}}+\frac{11 \pi 
^2}{4}-\frac{67}{6} \; .
\end{eqnarray}

%%%%%%%%%%%%%%%%%%%%%%%%%%%%%%%%%%%%%%%%%%%%%%%%%%%%%%%%%%%%%%%%%%%%%%%%%%%%%
%%%%%%%%%%%%%%%%%%%%%%%%%%%%%%%%%    References    %%%%%%%%%%%%%%%%%%%%%%%%%%
%%%%%%%%%%%%%%%%%%%%%%%%%%%%%%%%%%%%%%%%%%%%%%%%%%%%%%%%%%%%%%%%%%%%%%%%%%%%%


\begin{thebibliography}{99}


\bibitem{aur} 
P.~Aurenche, M.~Fontannaz, J.~P.~Guillet, B.~A.~Kniehl and M.~Werlen,
  %``Large p(T) inclusive pi0 cross-sections and next-to-leading-order QCD
  %predictions,''
  Eur.\ Phys.\ J.\  C {\bf 13}, 347 (2000)
  [arXiv:hep-ph/9910252].
  %%CITATION = EPHJA,C13,347;%%

\bibitem{apan}U.~Baur {\it et al.},
  %``Report of the working group on photon and weak boson production,''
  arXiv:hep-ph/0005226.
  %%CITATION = HEP-PH/0005226;%%

\bibitem{boursoff} C.~Bourrely and J.~Soffer,
 %``Do we understand the single-spin asymmetry for pi0 inclusive production  in
 %p p collisions?,''
  Eur.\ Phys.\ J.\  C {\bf 36}, 371 (2004)
  [arXiv:hep-ph/0311110].
  %%CITATION = EPHJA,C36,371;%%

\bibitem{kkp1}
  B.~A.~Kniehl, G.~Kramer and B.~P\"{o}tter,
  %``Testing the universality of fragmentation functions,''
  Nucl.\ Phys.\  B {\bf 597}, 337 (2001)
  [arXiv:hep-ph/0011155].
  %%CITATION = NUPHA,B597,337;%%

\bibitem{ddfwv}
  D.~de Florian and W.~Vogelsang,
  %``Threshold resummation for the inclusive hadron cross-section in p p
  %collisions,''
  Phys.\ Rev.\  D {\bf 71}, 114004 (2005)
  [arXiv:hep-ph/0501258].
  %%CITATION = PHRVA,D71,114004;%%

\bibitem{phenix} 
A.~Adare {\it et al.}  [PHENIX Collaboration],
%``Inclusive cross section and double helicity asymmetry for \pi^0 
%production in p+p collisions at sqrt(s)=200 GeV: Implications for the  
%polarized gluon distribution in the proton,''  
Phys.\ Rev.\  D {\bf 76}, 051106 (2007) [arXiv:0704.3599 [hep-ex]];
%A.~Adare {\it et al.}  [PHENIX Collaboration],
%``Inclusive cross section and double helicity asymmetry for pi^0 production 
%in p+p collisions at sqrt(s) = 62.4 GeV,''  
Phys.\ Rev.\  D {\bf 79}, 012003 (2009)  [arXiv:0810.0701 [hep-ex]].

\bibitem{star} J.~Adams {\it et al.}  [STAR Collaboration], 
%``Forward neutral pion production in p+p and d+Au collisions at  
% s(NN)**(1/2)
%= 200-GeV,'' 
Phys.\ Rev.\ Lett.\  {\bf 97}, 152302 (2006) [arXiv:nucl-ex/0602011].

\bibitem{brahms} I.~Arsene {\it et al.}  [BRAHMS Collaboration],  
%``Production of Mesons and Baryons at High Rapidity and High Pt in  
%Proton-Proton Collisions at sqrt(s) = 200 GeV,''  
Phys.\ Rev.\ Lett.\  {\bf 98}, 252001 (2007)  [arXiv:hep-ex/0701041].

\bibitem{na24} C.~De Marzo {\it et al.} [NA24 Collaboration],  
%``Measurement of the production of high mass gamma gamma, pi0 pi0, 
% and gamma 
%pi0 pairs in pi- p, pi+ p, and p p collisions at 300-GeV/c,''
Phys.\ Rev.\  D {\bf 42}, 748 (1990).

\bibitem{e711}
  H.~B.~White {\it et al.} [E711 Collaboration],
  %``Massive hadron pair production by 800-geV/c protons on nuclear targets,''
  Phys.\ Rev.\  D {\bf 48}, 3996 (1993);
  %%CITATION = PHRVA,D48,3996;%%
H.~B.~White, {\it A Study of angular dependence in parton-parton 
scattering from massive hadron pair production}, PhD Thesis, 
FERMILAB-THESIS-1991-39, FSU-HEP-910722, UMI-92-02321, 1991. 

\bibitem{e706}
M.~Begel [E706 Collaboration],
{\it Production of high mass pairs of direct photons and neutral mesons in a
Tevatron fixed target experiment}, PhD Thesis, FERMILAB-THESIS-1999-05, 
UMI-99-60725, 1999; see also: 
L.~Apanasevich {\it et al.}  [E706 Collaboration],  
%``Evidence for parton $k_{T}$ effects in high $p_{T}$ particle production,'' 
Phys.\ Rev.\ Lett.\  {\bf 81}, 2642 (1998) [arXiv:hep-ex/9711017].

\bibitem{ccor}
  A.~L.~S.~Angelis {\it et al.}  [CCOR Collaboration],
  %``RESULTS ON CORRELATIONS AND JETS IN HIGH TRANSVERSE MOMENTUM $p p$
  %COLLISIONS AT THE CERN ISR,''
Nucl.\ Phys.\ B {\bf 209}, 284 (1982). 

\bibitem{Chiappetta:1996wp}
 P.~Chiappetta, R.~Fergani and J.~P.~Guillet,
%``Production Of Two Large P(T) Hadrons From Hadronic Collisions,''
Z.\ Phys.\  C {\bf 69}, 443 (1996).

\bibitem{Owens:2001rr}
  J.~F.~Owens,
  %``A next-to-leading-order study of dihadron production,''
  Phys.\ Rev.\  D {\bf 65}, 034011 (2002)
  [arXiv:hep-ph/0110036].

\bibitem{Binoth:2002wa} 
T.~Binoth, J.~P.~Guillet, E.~Pilon and M.~Werlen,
%``A next-to-leading order study of pion-pair production and comparison with  
%E706 data,''  
Eur.\ Phys.\ J.\  C {\bf 24}, 245 (2002); 
%``A Next-to-leading order study of photon pion and pion pair 
% hadro production 
%in the light of the Higgs boson search at the LHC,'' 
Eur.\ Phys.\ J.\ direct C {\bf 4}, 7 (2002)  [arXiv:hep-ph/0203064].

\bibitem{dyresum} G.~Sterman, Nucl.\ Phys.\ B {\bf 281}, 310 (1987);
S.~Catani and L.~Trentadue, Nucl.\ Phys.\ B {\bf 327}, 323 (1989);
Nucl.\ Phys.\ B {\bf 353}, 183 (1991).

\bibitem{KS} 
N.~Kidonakis and G.~Sterman,
%``Resummation for QCD hard scattering,''
Nucl.\ Phys.\ B {\bf 505}, 321 (1997)
[arXiv:hep-ph/9705234].

\bibitem{BCMN}  R.~Bonciani, S.~Catani, M.~L.~Mangano and P.~Nason,
  %``Sudakov resummation of multiparton QCD cross sections,''
  Phys.\ Lett.\  B {\bf 575}, 268 (2003)
  [arXiv:hep-ph/0307035].
  %%CITATION = PHLTA,B575,268;%%

\bibitem{dyresum2} 
%
%\cite{Idilbi:2006dg}
%\bibitem{Idilbi:2006dg}
A.~Idilbi, X.~d.~Ji and F.~Yuan,
%``Resummation of Threshold Logarithms in Effective Field Theory For DIS,
%Drell-Yan and Higgs Production,''
Nucl.\ Phys.\  B {\bf 753}, 42 (2006) [arXiv:hep-ph/0605068];
%%CITATION = NUPHA,B753,42;%%
%
%\cite{Becher:2007ty}
%\bibitem{Becher:2007ty}
T.~Becher, M.~Neubert and G.~Xu,
%``Dynamical Threshold Enhancement and Resummation in Drell-Yan
%Production,''
JHEP {\bf 0807}, 030 (2008) [arXiv:0710.0680 [hep-ph]].
%%CITATION = JHEPA,0807,030;%%

\bibitem{KOS} N.~Kidonakis, G.~Oderda and G.~Sterman,
Nucl.\ Phys.\ B {\bf 525}, 299 (1998) [arXiv:hep-ph/9801268];
Nucl.\ Phys.\ B {\bf 531}, 365 (1998) [arXiv:hep-ph/9803241].

\bibitem{KO1}  
  N.~Kidonakis and J.~F.~Owens,
  %``Effects of higher-order threshold corrections in high-E(T) jet
  %production,''
  Phys.\ Rev.\  D {\bf 63}, 054019 (2001)
  [arXiv:hep-ph/0007268].
  %%CITATION = PHRVA,D63,054019;%%

\bibitem{Laenen:1992ey}
  E.~Laenen and G.~Sterman,
  %``Resummation for Drell-Yan differential distributions,''
  in {\it The Fermilab Meeting, DPF 1992} (World Scientific, 
 Singapore, 1993) Vol.\ 1, p.\ 987.
  %%CITATION = ITP-SB-92-69;%%
  
\bibitem{Sterman:2000pt}
  G.~Sterman and W.~Vogelsang,
  %``Threshold resummation and rapidity dependence,''
  JHEP {\bf 0102}, 016 (2001)
  [arXiv:hep-ph/0011289].
  %%CITATION = JHEPA,0102,016;%%

%\cite{Mukherjee:2006uu}
\bibitem{Mukherjee:2006uu}
  A.~Mukherjee and W.~Vogelsang,
  %``Threshold resummation for W-boson production at RHIC,''
  Phys.\ Rev.\  D {\bf 73}, 074005 (2006)
  [arXiv:hep-ph/0601162].
  %%CITATION = PHRVA,D73,074005;%%

\bibitem{cddf} G.~Bozzi, S.~Catani, D.~de Florian and M.~Grazzini,
   %``Higgs boson production at the LHC: Transverse-momentum resummation and
   %rapidity dependence,''
   Nucl.\ Phys.\  B {\bf 791}, 1 (2008) [arXiv:0705.3887 [hep-ph]].
  
%\cite{Sterman:2006hu}
\bibitem{Sterman:2006hu}
  G.~Sterman and W.~Vogelsang,
  %``Crossed threshold resummation,''
  Phys.\ Rev.\  D {\bf 74}, 114002 (2006)
  [arXiv:hep-ph/0606211].
  %%CITATION = PHRVA,D74,114002;%%
  
\bibitem{Cacciari:2001cw} 
M.~Cacciari and S.~Catani, 
%``Soft-Gluon Resummation for the Fragmentation of Light and Heavy Quarks at
%Large x,''  
Nucl.\ Phys.\  B {\bf 617}, 253 (2001) [arXiv:hep-ph/0107138].

\bibitem{KT} J.~Kodaira and L.~Trentadue, 
%%``Summing Soft Emission In QCD,''
Phys.\ Lett.\ B {\bf 112}, 66 (1982); Phys.\ Lett.\ B {\bf 123}, 
335 (1983);\\ S.~Catani, E.~D'Emilio and L.~Trentadue,
%%``THE GLUON FORM-FACTOR TO HIGHER ORDERS: GLUON GLUON ANNIHILATION AT SMALL
  %Q-TRANSVERSE,'
Phys.\ Lett.\ B {\bf 211}, 335 (1988).

\bibitem{top} L.~G.~Almeida, G.~Sterman and W.~Vogelsang,
%``Threshold Resummation for the Top Quark Charge Asymmetry,''
Phys.\ Rev.\  D {\bf 78}, 014008 (2008) [arXiv:0805.1885 [hep-ph]].

\bibitem{msj} M.~Sjodahl,
  %``Color structure for soft gluon resummation - a general recipe,''
  JHEP {\bf 0909}, 087 (2009)
  [arXiv:0906.1121 [hep-ph]].

\bibitem{Kidonakis:2001nj}
N.~Kidonakis, E.~Laenen, S.~Moch and R.~Vogt,
%``Sudakov resummation and finite order expansions of heavy quark
%hadroproduction cross sections,''
Phys.\ Rev.\  D {\bf 64}, 114001 (2001) [arXiv:hep-ph/0105041].

\bibitem{color} for related studies in heavy flavor production, see:
R.~Bonciani, S.~Catani, M.~L.~Mangano and P.~Nason,
%``NLL resummation of the heavy-quark hadroproduction cross-section,''
Nucl.\ Phys.\  B {\bf 529}, 424 (1998) [Erratum-ibid.\  B {\bf 803}, 
234 (2008)] [arXiv:hep-ph/9801375]; S.~Moch and P.~Uwer,
%``Theoretical status and prospects for top-quark pair production at hadron
%colliders,''
Phys.\ Rev.\  D {\bf 78}, 034003 (2008) [arXiv:0804.1476 [hep-ph]];
M.~Czakon and A.~Mitov,
%``On the Soft-Gluon Resummation in Top Quark Pair Production at Hadron
%Colliders,''
arXiv:0812.0353 [hep-ph].

\bibitem{KSV} A.~Kulesza, G.~Sterman and W.~Vogelsang,
  %``Joint resummation in electroweak boson production,''
  Phys.\ Rev.\  D {\bf 66}, 014011 (2002)
  [arXiv:hep-ph/0202251].

\bibitem{Catani} S.~Catani, D.~de Florian and M.~Grazzini,
%``Higgs production in hadron collisions: Soft and virtual QCD corrections  
%at NNLO,'' 
JHEP {\bf 0105}, 025 (2001) [arXiv:hep-ph/0102227]; see also:
M.~Kr\"{a}mer, E.~Laenen and M.~Spira, 
%``Soft gluon radiation in Higgs boson production at the LHC,'' 
Nucl.\ Phys.\  B {\bf 511}, 523 (1998) [arXiv:hep-ph/9611272];
R.~V.~Harlander and W.~B.~Kilgore,
%``Soft and virtual corrections to p p --> H + X at NNLO,''  
Phys.\ Rev.\  D {\bf 64}, 013015 (2001)  [arXiv:hep-ph/0102241].

\bibitem{Catani:1996yz} S.~Catani, M.~L.~Mangano, P.~Nason and L.~Trentadue,
  %``The Resummation of Soft Gluon in Hadronic Collisions,''
  Nucl.\ Phys.\  B {\bf 478}, 273 (1996)
  [arXiv:hep-ph/9604351].
  %%CITATION = NUPHA,B478,273;%%

\bibitem{cteq6} W.~K.~Tung, H.~L.~Lai, A.~Belyaev, J.~Pumplin, 
D.~Stump and C.~P.~Yuan, 
%``Heavy quark mass effects in deep inelastic scattering and global QCD
%analysis,''  
JHEP {\bf 0702}, 053 (2007) [arXiv:hep-ph/0611254].

\bibitem{DSS} D.~de Florian, R.~Sassot and M.~Stratmann, 
%``Global analysis of fragmentation functions for pions and kaons and their
%uncertainties,''  
Phys.\ Rev.\  D {\bf 75}, 114010 (2007) 
[arXiv:hep-ph/0703242].

\bibitem{akk} S.~Albino, B.~A.~Kniehl and G.~Kramer,
  %``AKK Update: Improvements from New Theoretical Input and Experimental
  %Data,''
Nucl.\ Phys.\  B {\bf 803}, 42 (2008) [arXiv:0803.2768 [hep-ph]].

\bibitem{Shimizu:2005fp}
  H.~Shimizu, G.~Sterman, W.~Vogelsang and H.~Yokoya,
  %``Dilepton production near partonic threshold in transversely polarized
  %proton-antiproton collisions,''
  Phys.\ Rev.\  D {\bf 71}, 114007 (2005)
  [arXiv:hep-ph/0503270].

%
\bibitem{eric} T.~O.~Eynck, E.~Laenen and L.~Magnea,  
%``Exponentiation of the Drell-Yan cross section near partonic threshold  in
%the DIS and MS-bar schemes,''  JHEP {\bf 0306}, 057 (2003)
[arXiv:hep-ph/0305179];
E.~Laenen and L.~Magnea,  
%``Threshold resummation for electroweak annihilation from DIS data,'' 
Phys.\ Lett.\  B {\bf 632}, 270 (2006) [arXiv:hep-ph/0508284];
%\cite{Laenen:2008gt}
%\bibitem{Laenen:2008gt}
E.~Laenen, G.~Stavenga and C.~D.~White,
%``Path integral approach to eikonal and next-to-eikonal exponentiation,''
JHEP {\bf 0903}, 054 (2009) [arXiv:0811.2067 [hep-ph]]; 
%%CITATION = JHEPA,0903,054;%%
%
%\cite{Laenen:2008ux}
%\bibitem{Laenen:2008ux}
E.~Laenen, L.~Magnea and G.~Stavenga,
%``On next-to-eikonal corrections to threshold resummation for the Drell-Yan 
%and DIS cross sections,''  
Phys.\ Lett.\  B {\bf 669}, 173 (2008) [arXiv:0807.4412 [hep-ph]].
%%CITATION = PHLTA,B669,173;%%

\bibitem{mjt} M.~J.~Tannenbaum,
%``From Bjorken scaling to pQCD: Experimental techniques from p - p collisions
%of the 1970's with application to Au + Au collisions at RHIC,''
Nucl.\ Phys.\  A {\bf 749}, 219 (2005) [arXiv:nucl-ex/0412004].

\bibitem{twoloopad}
%\cite{Aybat:2006mz}
%\bibitem{Aybat:2006mz}  
S.~M.~Aybat, L.~J.~Dixon and G.~Sterman,
%``The two-loop soft anomalous dimension matrix and resummation at 
%next-to-next-to leading pole,'' 
Phys.\ Rev.\  D {\bf 74}, 074004 (2006)  [arXiv:hep-ph/0607309];  
%%CITATION = PHRVA,D74,074004;%%
%%\cite{Aybat:2006wq}
%\bibitem{Aybat:2006wq}  
%S.~M.~Aybat, L.~J.~Dixon and G.~Sterman, 
%``The two-loop anomalous dimension matrix for soft gluon exchange,''  
Phys.\ Rev.\ Lett.\  {\bf 97}, 072001 (2006)  [arXiv:hep-ph/0606254];  
%%CITATION = PRLTA,97,072001;%%
%
%\cite{Gardi:2009qi}
%\bibitem{Gardi:2009qi}  
E.~Gardi and L.~Magnea,  
%``Factorization constraints for soft anomalous dimensions in QCD scattering  
%amplitudes,''  
JHEP {\bf 0903}, 079 (2009) [arXiv:0901.1091 [hep-ph]];
%%CITATION = JHEPA,0903,079;%%
%
%\cite{Becher:2009qa}
%\bibitem{Becher:2009qa} 
T.~Becher and M.~Neubert,  
%``On the Structure of Infrared Singularities of Gauge-Theory Amplitudes,''
JHEP {\bf 0906}, 081 (2009) [arXiv:0903.1126 [hep-ph]].  
%%CITATION = JHEPA,0906,081;%%

\bibitem{WvN} W.~L.~van Neerven,  
%``Dimensional Regularization Of Mass And Infrared Singularities In Two Loop 
%On-Shell Vertex Functions,'' 
Nucl.\ Phys.\  B {\bf 268}, 453 (1986).


\end{thebibliography}
\end{document}